\documentclass{aa}  

\usepackage{graphicx}
\graphicspath{{.}{figures/}}
\usepackage{txfonts}

\usepackage[bookmarks=false]{hyperref}

\usepackage{gensymb}  
\begin{document}

   \title{Revealing the structure of the outer disks of Be stars\thanks{Based on observations from the Karl J. Jansky Very Large Array collected via programme 10B-143, on observations from APEX collected via CONICYT programmes C-092.F-9708A-2013 and C-095.F-9709A-2015 and on observations from CARMA collected via programme c1100-2013a.}}


   \author{R.~Klement
          \inst{1,2}
          \and
          A.~C.~Carciofi\inst{3}
        \and
        T.~Rivinius\inst{1}
        \and
        L.~D.~Matthews\inst{4}
        \and
        R.~G.~Vieira\inst{3}
        \and
        R.~Ignace\inst{5}
        \and
        J.~E.~Bjorkman\inst{6}  
        \and
        B.~C.~Mota\inst{3}
        \and
        D.~M.~Faes\inst{3}
        \and
        A.~D.~Bratcher\inst{6}
        \and
        M.~Cur\'{e}\inst{7}
        \and
        S.~\v{S}tefl\thanks{Deceased}
}

  \institute{   European Southern Observatory, Alonso de Córdova 3107, Vitacura, Casilla 19001, Santiago, Chile\\
              \email{robertklement@gmail.com}
        \and
Astronomical Institute of Charles University, Charles University,
              V Hole\v sovi\v ck\'ach 2, 180 00  Prague 8
        \and
        Instituto de Astronomia, Geof\'isica e Ci\^encias Atmosf\'ericas, Universidade de S\~ao Paulo, Rua do Mat\~ao 1226, Cidade Universit\'aria, 05508-090, S\~ao Paulo, SP, Brazil
        \and
        MIT Haystack Observatory, off Route 40, Westford MA 01886, USA
        \and
        Department of Physics \& Astronomy, East Tennessee State University, Johnson City, TN 37614, USA
        \and
Ritter Observatory, Department of Physics \& Astronomy, University of Toledo, Toledo, OH 43606, USA
        \and
        Instituto de F\' isica y Astronom\' ia, Facultad de Ciencias, Universidad de Valpara\' iso, Casilla 5030, Valpara\' iso, Chile
}

   \date{}

 
  \abstract
   {The structure of the inner parts of Be star disks ($\lesssim 20$ stellar radii) is well explained by the viscous decretion disk (VDD) model, which is able to reproduce the observable properties of most of the objects studied so far. The outer parts, on the other hand, are not observationally well-explored, as they are observable only at radio wavelengths. A steepening of the spectral slope somewhere between infrared and radio wavelengths was reported for several Be stars that were previously detected in the radio, but a convincing physical explanation for this trend has not yet been provided.}
   {We test the VDD model predictions for the extended parts of a sample of six Be disks that have been observed in the radio to address the question of whether the observed turndown in the spectral energy distribution (SED) can be explained in the framework of the VDD model, including recent theoretical development for truncated Be disks in binary systems.}
   {We combine new multi-wavelength radio observations from the Karl.~G.~Jansky Very Large Array (JVLA) and Atacama Pathfinder Experiment (APEX) with previously published radio data and archival SED measurements at ultraviolet, visual, and infrared wavelengths. The density structure of the disks, including their outer parts, is constrained by radiative transfer modeling of the observed spectrum using VDD model predictions. In the VDD model we include the presumed effects of possible tidal influence from faint binary companions.  }
   {For 5 out of 6 studied stars, the observed SED shows strong signs of SED turndown between far-IR and radio wavelengths. A VDD model that extends to large distances closely reproduces the observed SEDs up to far IR wavelengths, but fails to reproduce the radio SED. Using a truncated VDD model improves the fit, leading to a successful explanation of the SED turndown observed for the stars in our sample. The slope of the observed SEDs in the radio is however not well reproduced by disks that are simply cut off at a certain distance. Rather, some matter seems to extend beyond the truncation radius, where it still contributes to the observed SEDs, making the spectral slope in the radio shallower. This finding is in agreement with our current understanding of binary truncation from hydrodynamical simulations, in which the disk does extend past the truncation radius. Therefore, the most probable cause for the SED turndown is the presence of binary companions that remain undetected for most of our sources. }
{}


   \maketitle
%

\section{Introduction}

Be stars offer unique possibilities for studying circumstellar disks. Because they are common among the bright, nearby stars -- 17\% of B-type stars are Be stars \citep{1997A&A...318..443Z} -- their disks are among typical targets for modern optical/near-infrared interferometers that have resolved them at the milliarcsec level. As a result, the structure of the inner parts of the disks is now well understood in the framework of the viscous decretion disk (VDD) model, first proposed by \citet{lee} and further developed by, for example, \citet{bjorkman1997}, \citet{porter1999}, \citet{okazaki}, and \citet{bjorkman_carciofi_2005}. The central stars rotate close to break-up velocities, and an uncertain mechanism -- the so-called Be phenomenon -- acts in addition to rotation, leading to episodic or continuous mass ejection from the stellar equator. Subsequently, outflowing, ionised, purely gaseous disks, rotating in a nearly Keplerian way are formed \citep[see][for a recent review]{review}. In the VDD model, it is the turbulent viscosity that is responsible for the transport of the angular momentum outwards, and therefore for the growth of the disk. 

In the last decades, observational techniques such as spectroscopy, linear polarimetry, and optical/near-IR interferometry have been used to constrain the physical structure of VDDs. Combining polarimetric and interferometric measurements, the disk-like structure of the circumstellar matter was unambiguosly confirmed \citep{quir} and the rotation law was subsequently confirmed to be nearly Keplerian \citep{meilland2007,wheel,kraus}. However, it is important to note that these results apply to the inner parts of VDDs only \citep[$\lesssim 20$ stellar radii,][]{vieira}, that are responsible for the bulk of the disk emission in the optical and infrared (IR), as well as for the Balmer lines \citep{carciofi2011}. At larger radii, the disks are detectable only at radio wavelengths. 

In this work, we are interested in the overall density structure of VDDs, focusing on the parts outwards of $\sim$20 stellar radii. We constrain the density structure by compiling the observed spectral energy distribution (SED) from ultraviolet (UV) up to radio wavelengths. Due to the nature of the disk continuum emission -- bound-free and free-free emission from the ionized hydrogen in the disk -- fluxes at longer wavelengths originate from progressively larger areas of the disk. At the UV wavelengths, the SED usually consists of the stellar photospheric flux only, although in shell stars (Be stars seen close to edge-on), the UV photospheric spectrum can be dimmed by the surrounding disk and also strongly blanketed by the disk line absorption (mostly Fe). At longer wavelengths, the disk contribution to the SED becomes more significant until it dominates the SED in the mid-IR to radio domains. This is well explained in terms of the gas in the disk forming an optically thick pseudo-photosphere, whose size grows with increasing wavelength \citep[see][for more details on the pseudo-photosphere concept applied to the continuum emission of Be disks]{vieira}. 

Radio observations longwards of sub-mm wavelengths are of special interest for studying the outer parts of VDDs, as the disk pseudo-photospheres are larger at these wavelengths. In this work, we revisit the Be stars that were detected by the Very Large Array (VLA) observations at 2~cm by \citet[][six detected stars]{taylor} and subsequently observed at mm wavelengths using the James Clerk Maxwell Telescope (JCMT) by \citet{waters}. The main outcome of the radio measurements was the discovery of an SED turndown, that is, a steepening of the spectral slope somewhere between mid-IR (10 -- 60 $\mu$m, as measured by the IRAS satellite) and radio wavelengths, in the studied stars. Using a simple wedge-shaped disk model seen pole-on, \citet{waters} were able to reproduce the observed SEDs with the disks having power-law density structure in the form $\rho \propto r^{-n}$, with the exponent $n$ ranging from 2.0 to 3.5. However, to reproduce the SED turndowns, disks truncated at a certain radius (ranging from 26 to 108 stellar radii) had to be used. Although the VDD model was not yet established at that time, it was already suspected that Be envelopes have disk-like shapes that originate from the central star and flow outwards. If the disks are dominated by rotation rather than outflow (as in the VDD model), it would follow that the most likely reason for the disk truncation is the presence of an (unseen) binary companion. However, because \citet{waters} assumed wind-like outflows rather than disks dominated by rotation, they found truncation to be an unlikely explanation for the SED turndown, favoring other possibilities, such as changes in the disk geometry or additional acceleration of the disk material in the outer disk.

In the last decade or so, the VDD model has been firmly established as the framework in which most of the observable properties of Be disks can be explained (at least in their inner parts). Successful VDD reproductions of multi-technique and multi-wavelength observations of individual objects include, for example, $\zeta$ Tauri \citep{ztau2}, $\beta$ CMi \citep{klement,klement_aspc} and 48~Lib \citep{silaj}. This allows us to interpret the radio observations on a firm physical basis and therefore distinguish between the suggested scenarios to explain the SED turndown.  

The goal of the present work is to carefully model the compiled SEDs of the Be stars detected in the radio by \citet{taylor} using the VDD model implemented in the Monte Carlo radiative transfer code {\ttfamily HDUST} \citep{hdust1,hdust2}. For four of these stars, we have obtained new multi-band measurements in the cm domain, allowing us to study the properties of the SED turndown in detail. With the results of this study we address the question of the origin of the SED turndown, that is, whether it can be explained in the framework of the VDD model, which may be truncated by a binary companion, or if further physical ingredients in the model are needed. 

In the following section we focus on the theoretical predictions concerning the structure of the outer disks of Be stars. In Sect.~3 we describe the observations used, and in Sect. 4 we explain the VDD model and the modeling procedure. The following section details the results for the six individual objects and the conclusions follow.

\begin{table*}[t]
 \caption[]{\label{timeline}Details of the IR, sub-mm, and radio observations.}
\centering
\begin{tabular}{lccc}
 \hline \hline
  Mission/Telescope & Epoch of observations  & $\lambda_\text{central}$/phot. band & Reference \\
\hline
IRAS    & 1 -- 11/1983     & 12, 25, 60, 100\,$\mu$m &  \citet{helou} \\
VLA & 12/1987 -- 9/1988, 2/1990, 8/1991     & 2, 3.6, 6\,cm & \citet{taylor1987,taylor},\\
& & & \citet{apparao},\\
& & & \citet{radio_var}, \\
& & & \citet{dougherty_nature}\\
RAO/NOAO/MKO & 4/1988 -- 3/1991   & $JHKLMN$ bands & \citet{dougherty}\\
JCMT & 8/1988, 8/1989   & 0.76, 1.1\,mm & \citet{waters1989,waters}\\
IRAM & 3/1991, 12/1992   & 1.2\,mm & \citet{wendker}\\
MSX & 1996 -- 1997     & 4.29, 4.35, 8.88, 12.13, 14.65, 21.3\,$\mu$m & \citet{msx}\\
AKARI & 5/2006 -- 8/2007    & 9, 18, 65, 90, 140, 160\,$\mu$m & \citet{ishihara}\\
WISE & 1/2010 -- 11/2010  & 3.3, 4.6, 12 and 22\,$\mu$m & \citet{wise}\\
JVLA & 10/2010    & 0.7, 1.3, 3.5 and 6.0\,cm & this work\\
CARMA & 9/2013 & 3\,mm & \citet{klement} \\
APEX/LABOCA & 9/2013, 7--8/2015  & 0.87\,mm & this work\\
\hline
\end{tabular}
\end{table*}

\section{The outer parts of Be star disks}

In an isolated VDD, the outflow velocity in the inner disk is highly subsonic and grows linearly with the distance from the central star \citep{okazaki}. At a certain point, sometimes called the critical or photo-evaporation radius, the outflow velocity reaches the local sound speed, which marks the transition between a subsonic inner part and a supersonic outer part of the disk. Outwards from this point, it is no longer viscosity but rather the gas pressure that drives the outflow, and the disk becomes angular momentum conserving. This transition may offer an explanation for the SED turndown, as the density decreases much more rapidly past the transonic part. However, the transition is only expected to occur at very large distances from the central star. An approximate relation for the critical radius $R_\text{c}$ was derived by \citet{krticka} for isothermal disks:
\begin{equation}
\label{krticka}
\frac{R_\text{c}}{R_\mathrm{e}} = \frac{3}{10}\left(\frac{v_\text{orb}}{c_\text{s}}\right)^2,
\end{equation}
where $R_\mathrm{e}$ is the stellar equatorial radius, $v_\text{orb}$ is the disk orbital speed above the stellar equator and $c_\text{s}$ is the speed of sound. The typical values of $R_\text{c}$ are approximately 350 and 430\,$R_*$ for spectral types B0 and B9, respectively. According to the pseudo-photosphere model of \citet{vieira}, for typical disk densities, the pseudo-photosphere radius attains such large values only at wavelengths longer than 10\,cm. Unless the disk is extremely dense, this means that only wavelengths longwards of 10\,cm can probe the transonic regions. So far, no detections of Be disk emission at these wavelengths have been reported. It follows that the SED turndown is likely unrelated to the existence of a transonic regime in Be disks.

The only known physical mechanism that can truncate the disk at a closer distance to the central star is the tidal influence from an orbiting secondary companion. The effects of such a scenario on the VDD of the primary have  recently been explored by smoothed particle hydrodynamics (SPH) simulations, under the assumption that the orbit of the secondary is aligned with the disk \citep{panoglou}. For circular orbits, the influence of the secondary has two important effects on the Be disk. The first is the truncation of the disk at a distance close to the 3:1 resonance radius, which is much smaller than the orbital separation. Outwards of the truncation radius, the density starts to decrease much more rapidly, but these parts can still contribute to the emergent spectrum. The second is the so-called accumulation effect, which causes the parts of the disk inwards of the truncation radius to have a slower density fall-off (i.e., a smaller radial power law exponent). The accumulation effect has a larger impact with decreasing orbital period, decreasing viscosity, and increasing binary mass ratio. Considering the possibility that the orbit of the secondary is misaligned with respect to the equatorial plane of the disk, both the truncation and accumulation effects become much weaker \citep{cyr}.

The incidence of binarity among Be stars remains an important open issue. The presence of a close binary companion was, in the past, believed to be the cause of the Be phenomenon, although that is not likely the case \citep{oudmaijer_parr}. However, the evolution of a mass-transferring binary may produce a low-mass subdwarf O or B star (sdO/sdB, the mass donor that was originally the more massive component and is now stripped of its outer layers) and a fast-spinning star (the mass and angular momentum gainer, that is now the primary star). It is expected that at least some Be stars may have formed in this fashion. Owing to their lower luminosity, sdO/sdB companions are difficult to detect and only five Be+sdO/sdB systems have been reported so far: $\varphi$~Per \citep{gies1998}, FY CMa \citep{peters2008}, 59~Cyg \citep{peters2013}, $o$~Pup \citep{koubsky_opup}, and HR~2142 \citep{hr2142}. They were revealed by periodic signatures in the UV spectra and by traveling emission components in the emission lines of the Be star caused by radiative interaction of the hot sdO/sdB star with the outer disk of the Be star. Other common companions of Be stars are neutron stars, which may emit X-rays as they accrete some of the matter from the disk of the Be primary. Other compact companions should include black holes and white dwarfs, although only one such system has been confirmed \citep[Be star + black hole system HD 215227;][]{casares}. The companions that are hardest to detect, though they are possibly common, are low-mass main-sequence stars.

Searching for binary companions using radial velocity (RV) shifts of spectral lines is complicated by their large rotational broadening. The exceptions from this rule are shell stars, which have narrow absorption lines in the centers of broad emission lines formed in the disk and thus orbiting companions are more easily detectable. Searches for binary companions of Be stars using speckle interferometry \citep{mason1997} and direct imaging with adaptive optics \citep{roberts2007, oudmaijer_parr} led to the conclusion that the incidence of binaries among Be stars is around 30\%. However, these results apply only for companions that are farther than 0.1 arcsec ($\sim 20$\,au for the most nearby targets), with a maximum magnitude difference of 10 mag in the $I$ and $K$-bands. That means that Be companions such as main sequence objects of spectral classes F and cooler or sdO/sdB stars are not detectable with these techniques. Such binary companions would remain invisible, but could, in principle, be revealed through their influence on the density structure of the disk of the primary. If prevalent, the truncation of Be disks by such companions could be the cause of the observed turndown in the SEDs. Considering additional indirect evidence, line emission in the IR Ca triplet has been suggested to indicate binarity in Be stars \citep[see, e.g., discussion by][]{2011JPhCS.328a2026K}. Also, in known Be binaries in a close orbit (e.g., $\gamma$~Cas), the H$\alpha$ line often shows no clear peaks or symmetry, which is thought to indicate disturbances in the disk due to the orbiting companion. 


\section{Observations}

The primary dataset used for the modeling is the SED measurements, namely the absolutely calibrated UV spectra and photometric measurements from visual to radio wavelengths. The secondary dataset contains visual spectra and polarimetry. The spectra were used to search for RV variations that could be assigned to the presence of an orbiting companion. The emission line profiles and the polarimetry were used to check for variability in the disk throughout the last decades.

The epochs, wavelengths, and references of the IR, sub-mm, and radio photometric measurements are given in Table~\ref{timeline}. The epochs of the observations span over three decades. Since Be disks are often highly variable, this may introduce errors in our solutions. Nevertheless, the data for each target were still combined into a single dataset, which then represents average properties of the disk over the last decades. For more details on the variability of individual targets, and how it affects the solution, we refer to Sect.~\ref{results}.

\onltab{
\begin{table*}[t]
  \begin{center}
    \caption{\label{vla_onl1}VLA Calibration Sources.}
      \begin{tabular}{l ccccc} \\
        \hline
\hline
      Source & $\alpha$(J2000.0) & $\delta$(J2000.0) & Date & $\nu_{0}$ (MHz) & Flux Density (Jy) \\
\hline
J0102+5824$^{a}$ & 01 02 45.7624 & 58 24 11.137 & 21-OCT-2010 &4.894 & 1.321$\pm$0.004\\
     & &                                  & 21-OCT-2010 & 5.022 & 1.347$\pm$0.004\\
     & &                                  & 30-OCT-2010 &8.394 & 1.965$\pm$0.006\\
     & &                                   & 30-OCT-2010 &8.522& 1.969$\pm$0.005\\
     & &                                  &31-OCT-2010 &8.394 & 1.937$\pm$0.003\\
     & &                                   & 31-OCT-2010 &8.522& 1.948$\pm$0.003\\
     & &                                  & 30-OCT-2010 & 22.394 &2.28$\pm$0.04 \\
     & &                                  & 30-OCT-2010 & 22.522 &2.26$\pm$0.03\\
    & &                                   & 31-OCT-2010 & 22.394 & 2.08$\pm$0.02\\
    & &                                   & 31-OCT-2010 & 22.522 & 2.09$\pm$0.01\\
    & &                                    & 30-OCT-2010 & 43.342 &2.03$\pm$0.10\\
    & &                                    & 30-OCT-2010 & 43.470 & 2.04$\pm$0.10\\
    & &                                    & 31-OCT-2010 & 43.342 & 1.82$\pm$0.04\\
    & &                                    & 31-OCT-2010 & 43.470 & 1.82$\pm$0.04\\
3C48$^{b}$ & 01 37 41.2994 & +33 09 35.132 & 11-OCT-2010 & 4.895 & 5.3854$^{*}$\\
      & &                                & 12-OCT-2010 & 4.895 & 5.3855$^{*}$\\
      & &                                & 21-OCT-2010 & 4.895 & 5.3857$^{*}$\\
      & &                                & 31-OCT-2010 & 4.895 & 5.3859$^{*}$\\
      & &                                & 15-OCT-2010 & 8.395 & 3.2074$^{*}$\\
      & &                                & 30-OCT-2010 & 8.395 & 3.2080$^{*}$\\
      & &                               & 31-OCT-2010 & 8.395 & 3.2081$^{*}$\\
      & &                               & 15-OCT-2010 &22.395 & 1.2422$^{*}$\\
      & &                               & 30-OCT-2010 &22.395 & 1.2429$^{*}$\\
      & &                               & 31-OCT-2010 &22.395 & 1.2430$^{*}$\\
      & &                               & 15-OCT-2010 &43.215 & 0.6739$^{*}$\\
      & &                               & 30-OCT-2010 &43.343 & 0.6728$^{*}$\\
      & &                               & 31-OCT-2010 &43.343 & 0.6728$^{*}$\\
J0349+4609$^{c}$ & 03 49 18.7416 & 46 09 59.658 & 31-OCT-2010 & 8.394 &0.5802$\pm$0.0005\\
           & &                               & 31-OCT-2010 & 8.522 &0.5772$\pm$0.0005\\
           & &                               & 31-OCT-2010 & 22.394& 0.410$\pm$0.002\\
           & &                                & 31-OCT-2010 & 22.522& 0.407$\pm$0.002 \\
           & &                                & 31-OCT-2010 & 43.342 &0.335$\pm$0.005\\
           & &                                & 31-OCT-2010 & 43.470&0.333$\pm$0.006\\
J0359+5057$^{c}$ & 03 59 29.7473 & 50 57 50.162 & 11-OCT-2010&4.894 &8.56$\pm$0.02\\
      & &                                    & 11-OCT-2010&5.022 &8.71$\pm$0.02\\       
J0403+2600$^{d}$ & 04 03 05.5860 & 26 00 01.503& 31-OCT-2010 &4.894 & 2.344$\pm$0.003\\
     & &                                    & 31-OCT-2010 & 5.150 & 2.357$\pm$0.004\\
     & &                                 & 15-OCT-2010 & 8.394 & 2.190$\pm$0.003\\ 
     & &                                 & 15-OCT-2010 & 8.650 & 2.187$\pm$0.003\\
     & &                                 & 15-OCT-2010 & 22.394& 1.55$\pm$0.01\\
     & &                                 & 15-OCT-2010 & 22.650 & 1.56$\pm$0.01\\
     & &                                 & 15-OCT-2010 & 43.214& 1.15$\pm$0.04 \\
     & &                                 & 15-OCT-2010 & 43.470 & 1.15$\pm$0.04\\
J2311+4543$^{e}$ & 23 11 47.4090 & 45 43 56.016 & 31-OCT-2010& 8.394&0.602$\pm$0.001 \\
    & &                                      & 31-OCT-2010& 8.650& 0.605$\pm$0.001\\
    & &                                      & 31-OCT-2010& 22.394& 0.525$\pm$0.006\\
    & &                                      & 31-OCT-2010& 22.650& 0.531$\pm$0.007\\
    & &                                      & 31-OCT-2010& 43.342& 0.43$\pm$0.01\\
    & &                                      & 31-OCT-2010& 43.598& 0.43$\pm$0.01\\
J2322+5057$^{e}$ & 23 22 25.9822 & 50 57 51.964 &12-OCT-2010 & 4.894 & 1.251$\pm$0.008\\
     & &                                     & 12-OCT-2010 & 4.894 & 1.245$\pm$0.008\\
\hline
    \end{tabular}
  \end{center}
\tablefoot{}
\tablefoottext{*}{Adopted flux density, based on \citet{perley}}
\tablefoottext{a}{Secondary gain calibrator for $\gamma$~Cas.}
\tablefoottext{b}{Absolute flux density and bandpass calibrator.}
\tablefoottext{c}{Secondary gain calibrator for $\psi$~Per.}
\tablefoottext{d}{Secondary gain calibrator for $\eta$~Tau.}
\tablefoottext{e}{Secondary gain calibrator for EW~Lac.}
For 3C48, the flux density $S_{\nu}$ as a function of frequency was taken to be ${\rm log}(S_{\nu}) = 1.3332 - 0.7665({\rm   log}(\nu)) - 0.1981({\rm log}(\nu))^{2} +0.0638({\rm log}(\nu))^3$, where $\nu_{\rm GHz}$ is the frequency expressed in GHz. 
\end{table*}
}

\onltab{
\begin{table*}[t]
  \begin{center}
    \caption{\label{vla_onl2}Deconvolved image characteristics for the new VLA Data.}
      \begin{tabular}{l cccc} \\
        \hline
\hline
Source & $\lambda$ [cm] & $\theta_{\rm FWHM}$ [arcsec] & PA [degrees]& rms [$\mu$Jy beam$^{-1}$]\\
\hline
$\gamma$~Cas & 6.0 & 5.4$\times$4.0 & $-59.3$ & 11.0 \\
$\gamma$~Cas$^{*}$ & 3.5 & 2.9$\times$2.3 & $-43.6$ & 16.0\\
$\gamma$~Cas$^{*}$ & 1.3 & 1.1$\times$0.89 & $-29.6$ & 56.4 \\
$\gamma$~Cas$^{*}$ & 0.7 & 0.63$\times$0.51 & $-33.7$ & 156\\

$\psi$~Per & 6.0 & 6.7$\times$4.7 & +88.7 & 12.3 \\
$\psi$~Per & 3.5 & 2.6$\times$2.3 & +9.8 & 21.9 \\
$\psi$~Per & 1.3 & 1.1$\times$1.0 & +12.4 & 83.5\\
$\psi$~Per & 0.7 & 0.67$\times$0.55 & +29.7 & 252 \\

EW~Lac & 6.0 & 6.3$\times$4.6 & $-89.6$ & 11.7\\
EW~Lac & 3.5 & 2.7$\times$2.2 & +35.1 & 23.4\\
EW~Lac & 1.3 & 1.2$\times$0.9 & +30.6 & 76.5\\
EW~Lac  & 0.7 & 0.60$\times$0.53 & +57.0 & 275 \\

$\eta$~Tau &  6.0 & 4.6$\times$4.1 & $-3.2$ & 9.75\\
$\eta$~Tau & 3.5 & 2.7$\times$2.2 & $-2.5$ & 23.9 \\
$\eta$~Tau & 1.3 & 1.0$\times$0.9 & +14.9 & 91.1 \\
$\eta$~Tau & 0.7 & 0.62$\times$0.52 & +56.9 & 297.5\\
\hline
\end{tabular}
\end{center}
\tablefoot{}
\tablefoottext{*}{For $\gamma$~Cas, observations from 2010 October 30 and 31 were combined to make the final maps.}
\end{table*}
}

\begin{table*}[t]
  \begin{center}
    \caption{\label{radio_obs_new}Newly acquired radio observations from JVLA (0.7, 1.3, 3.5 and 6.0\,cm).}
      \begin{tabular}{l ccccc} \\
        \hline
\hline
      Star & $F$ at 0.7\,cm [mJy] & $F$ at 1.3\,cm [mJy] & $F$ at 3.5\,cm [mJy] & $F$ at 6.0\,cm [mJy]\\
\hline
$\eta$ Tau                 & 2.09 $\pm$ 0.30 & 0.852 $\pm$ 0.090 & 0.237 $\pm$ 0.024 & 0.1432 $\pm$ 0.0098 \\
EW Lac                     & $<$0.83 (3$\sigma$) & 0.351 $\pm$ 0.077 & 0.133 $\pm$ 0.023 & 0.089 $\pm$ 0.012 \\
$\gamma$ Cas           & 1.65 $\pm$ 0.15 & 0.710 $\pm$ 0.057 & 0.201 $\pm$ 0.016 & 0.128 $\pm$ 0.011  \\
$\psi$ Per         & 3.47 $\pm$ 0.26 & 1.284 $\pm$ 0.083 & 0.435 $\pm$ 0.022 & 0.241 $\pm$ 0.012 \\
\hline
    \end{tabular}
  \end{center}
\end{table*}

\subsection{Ultraviolet spectra}

For the UV part of the spectrum, data from the International Ultra\-violet Explorer (IUE)\footnote{\url{https://archive.stsci.edu/iue/}} and the Wisconsin Ultra\-violet Photo-Polarimeter Experiment (WUPPE)\footnote{\url{https://archive.stsci.edu/wuppe/}} were used. For the IUE spectra, large aperture measurements were used in all cases due to the problems with absolute flux calibration of the small aperture measurements. When available, we preferred low-dispersion spectra. The data show very little or no temporal variation for all stars and were therefore averaged for modeling purposes.

\subsection{Visual and IR photometry}

To compile all the available visual and IR photometric measurements, we made use of the Virtual Observatory SED Analyser\footnote{\url{http://svo2.cab.inta-csic.es/theory/vosa/}} \citep[VOSA;][]{vosa} and the VO Spectral Analysis Tool\footnote{\url{http://esavo.esa.int/vospec/}} \citep[VOSpec;][]{vospec}. These tools search the available catalogs and compile the SED for a given object. The visual domain catalogs that VOSA and VOSpec searched and that we subsequently used in this study include those of \citet{ducati}, Hipparcos \citep{hipparcos}, \citet{ubvmerm} and Tycho-2 \citep{tycho}. For the IR domain, data from the space mission MSX \citep[MSX6C Infrared Point Source Catalog;][]{msx} was used. 

In the near-IR domain we made use of $JHKLMN$ magnitudes measured at three observatories: the University of Calgary Rothney Astrophysical Observatory (RAO) 1.5\,m telescope, the National Optical Astronomy Observatory (NOAO), Kitt Peak 1.3\,m telescope and the MKO 0.6\,m telescope \citep{dougherty}. In the mid-IR domain, we used available data from IRAS \citep{helou}, AKARI \citep[AKARI/IRC mid-IR all-sky Survey;][]{ishihara}, and WISE \citep[][]{wise} satellites. The catalog flux values were color-corrected in the same way as described by \citet{vieira2017}. For $\beta$~CMi we also used data from the Spitzer Space Telescope \citep[SST; ][]{spitzer}. IR measurements corresponding to upper limits were discarded. 

We note that for the star $\beta$~Mon~A, the visual and IR photometry actually corresponds to the combined photometry of the three components, all of which are Be stars of similar spectral types. This may introduce additional errors in our solution.

\subsection{Radio observations}

\subsubsection{Sub-millimeter observations}

For the star $\beta$ Mon A, we used a new APEX measurement from the bolometer camera LABOCA \citep{siringo} at the wavelength of 870 $\mu$m. The observed flux is $23.5 \pm 3.5$\,mJy. The separation between the A and B components is 7.1 arcsec. When centered on the A component, the B component falls inside the FWHM of the LABOCA beam, which is equal to $19.2 \pm 3$ arcsec. Although close to the edge of the FWHM, the B component may still slightly contribute to the observed flux.

For $\beta$~CMi, we also have an APEX/LABOCA measurement, that was originally published in \citet{klement}. However, we newly reduced the data using Crush\footnote{\url{http://www.submm.caltech.edu/~sharc/crush/index.html}} version 2.32-1 \citep{crush}, with the updated flux value for $\beta$~CMi being $38.6 \pm 3.1$\,mJy. This value remains consistent within the error bars with the value used previously. For a more detailed description of the APEX/LABOCA measurements, see Sect. 2.2 of \citet{klement}. For $\beta$~CMi, we also use the flux at 3\,mm as measured by CARMA \citep[see Table~1 of][]{klement}.

\subsubsection{Millimeter observations}

At mm wavelengths, we used the previously published JCMT measurements of \citet{waters1989,waters} and the IRAM measurements of \citet{wendker}. 

\subsubsection{Centimeter observations}

The previously published radio fluxes at cm wavelengths were taken from the VLA observations presented by \citet{taylor1987,taylor}, \citet{apparao}, \citet{radio_var} and \citet{dougherty_nature}. 

New radio observations of four stars were obtained with the Karl G. Jansky Very Large Array (JVLA) of the National Radio Astronomy Observatory in October, 2010 (Table~\ref{radio_obs_new}). Each star was observed at four different wavelengths (6.0~cm, 3.5~cm, 1.3~cm, and 0.7~cm) in the C configuration (baselines $\sim0.035 - 3.4$~km). For all four observing bands, the WIDAR correlator was configured using a setup that provided two sub-bands, each with a 128~MHz bandwidth, 64 spectral channels, and dual circular polarizations. Typical on-source integration times for each star were $\sim$65, 13, 4, and 3.5~minutes at 6.0~cm, 3.5~cm, 1.3~cm, and 0.7~cm, respectively, with the exception of $\gamma$~Cas, where a second observation of comparable duration was obtained at each of the three shortest wavelengths. Observations of each star were interspersed with observations of a neighboring bright point source to serve as a complex gain calibrator, and each session included an observation of 3C48  at the appropriate observing band(s) to serve as an absolute flux density and bandpass calibrator. Data reduction was performed using the Astronomical Image Processing System \citep[AIPS;][]{greisen} and followed standard procedures for continuum data obtained with the WIDAR correlator as described in the AIPS Cookbook\footnote{\tt http://www.aips.nrao.edu/cook.html}. The data were loaded into AIPS directly from the archival science data model (ASDM)  files using the  Obit software package \citep{cotton}. However, the default calibration (`CL') tables were  recreated  to  update the gain and  opacity information. The antenna positions were also updated to the best available values.

For all of the data obtained for this program it was necessary to perform an initial delay correction using a short ($\sim$1 minute) segment of data from the flux/bandpass calibrator. Following the excision of poor-quality data, the data weights were calibrated based on the system power measurements   from each antenna using the AIPS task {\sc TYAPL}. The absolute flux density scale was determined using the time-dependent 3C48 flux density values from \citet{perley}. Calibration of the bandpass and the frequency-independent portion of the complex gains was subsequently performed using standard techniques. At the two highest frequencies, the gain calibration was performed in two steps, first solving only for the phases using a solution interval of $\sim$10~seconds. Following the application of these corrections, a second iteration solved for both amplitude and phase.  The flux densities determined for each of the complex gain calibrators observed in the present study are summarized in Table~\ref{vla_onl1} (online only).

Imaging was performed using the AIPS task {\sc IMAGR} with robust +1 weighting of the visibilities. Table~\ref{vla_onl2} (online only) provides a summary of the resulting properties of the image and the synthesized beam. Flux densities for each of the target stars were then determined based on Gaussian fits to these images using the AIPS task {\sc JMFIT}). All of the stars were spatially unresolved, hence the measured peak flux density is taken to represent the total flux. 

\subsection{Visual spectroscopy}

Visual domain spectra for the program stars were  obtained from
the following archives:
\begin{itemize}
\item Unpublished {\sc Heros} data, see \citet{2005PAICz..93....1S}.

\item The ELODIE archive described by \citet{2004PASP..116..693M}.

\item The ESPaDOnS archive at CADC\footnote{\tt
http://www.cadc-ccda.hia-iha.nrc-cnrc.gc.ca/en/}. ESPaDOnS is
mounted at the CFHT on Hawaii and described by \citet{2006ASPC..358..362D}.

\item The TBL/Narval archive\footnote{\tt
tblegacy.bagn.obs-mip.fr/narval.html}. NARVAL is a copy of the ESPaDOnS instrument, adapted to the 2\,m TBL.

\item The BeSS database \citep{2011AJ....142..149N}. However,
data from this source were not used for velocity measurements;
only for equivalent width (EW) measurements and general trend descriptions.
\end{itemize}

\subsection{Visual polarimetry}

Visual spectropolarimetric measurements were taken from the archives of the HPOL spectropolarimeter\footnote{\url{http://archive.stsci.edu/hpol/}} mounted on the Pine Bluff Observatory (PBO) 36'' telescope \citep{2012AIPC.1429..226M}.


\section{SED modeling}

\subsection{Model description}

The adopted model of a Be star system consists of a combination of a fast rotating central star and a VDD.

The central star is approximated by a spheroidal, rotationally oblate shape that is gravity darkened according to the von Zeipel law \citep{vonzeipel}. Its surface is divided into a number of latitude bins, and each one is assigned with a model spectrum \citep{kurucz} corresponding to the local values of $T_\text{eff}$ and $\log{g}$.

The rotating central star is thus completely described by four independent parameters: mass $M$, polar radius $R_\text{p}$, luminosity $L$, and rotation rate $W$. The $W$ parameter is defined as $W = v_{\text{rot}}/v_{\text{orb}}$, where  $v_{\text{rot}}$ is the rotational velocity at the stellar surface and $v_{\text{orb}}$ is the Keplerian orbital velocity directly above the stellar surface. The gravity darkening exponent $\beta$ that appears in the classical form of the gravity darkening law, $T_{\text{eff}} \propto g^{\beta}$ \citep{vonzeipel}, is a function of $W$ as given by \citet{lara}. 

The VDD model is adopted in a simplified parametric form, in which the density structure follows a power law decrease in the radial direction and has a hydrostatic equilibrium structure in the vertical direction:
\begin{equation}
\rho(r,z) = \rho_0 \left(\frac{r}{R_\text{e}}\right)^{-n} \text{exp}\left(-\frac{z^2}{2H^2}\right),
\end{equation}
where $r$ and $z$ are respectively the radial and vertical cylindrical coordinates, $\rho_0$ is the density of the disk close to the stellar surface, $R_\text{e}$ is the equatorial radius of the rotationally deformed central star, $n$ is the density exponent, and  $H$ is the disk scale height. The disk flares with the following dependence of the disk scale height $H$ on the distance $r$ from the star: 
\begin{equation}
\label{sch}
H(r) = H_0 \left(\frac{r}{R_\text{e}}\right)^{3/2},
\end{equation}
where $H_0 = c_\text{s} v_\text{orb}^{-1} R_\text{e}$ is the pressure-supported scale height at the base of the disk, $c_\text{s} = [(k_\text{B} T_\text{k})/(\mu m_\text{a})]^{1/2}$ is the isothermal sound speed, $k_\text{B}$ is the Boltzmann constant, $T_\text{k}$ is the gas kinetic temperature, $\mu$ is the mean molecular weight, and $m_\text{a}$ is the atomic mass unit. $T_\text{k}$ is the parameter that determines the scale height, and we choose it to correspond to 60\% of the effective temperature at the stellar equator, following \citet{hdust1}. 

For $n=3.5$, this structure represents the subsonic parts of an isolated, isothermal VDD in a steady-state (when mass from the central star was fed to the disk at a constant rate for a sufficiently long time) and composed purely of hydrogen \citep{bjorkman_carciofi_2005}. However, there are effects that can alter the value of $n$. Local non-isothermalities in the disk can affect both the flaring exponent and the radial density fall-off \citep{hdust2}, while a non-steady mass feeding rate \citep{haubois1}, and the accumulation effect caused by a close orbiting companion \citep{panoglou} can also affect the density profile in quite complicated ways. For instance, \citet{haubois1} found that a forming disk (i.e., a disk being actively fed by the star but that has not yet reached a steady-state configuration) has a sharp density profile, which, if approximated by a power-law, results in $n$ larger than 3.5; conversely, a dissipating disk (i.e., a disk that is no longer being fed and is therefore accreting back to the central star) has a much flatter density exponent ($n \approx 3.0$). Density exponents smaller than 3.5 are also believed to ensue from the accumulation effect in Be binaries. In this regard, varying $n$ in our models allows us to accommodate these processes in the studied targets.

The only remaining free parameter of the disk is its outer radius $R_\text{out}$. If not set to a very high number corresponding to a steadily fed isolated disk, this parameter roughly corresponds to the truncation radius caused by the tidal influence of a possible binary companion. In this work, we adopt a sharp truncation, in which no material exists past $R_\text{out}$. Detailed hydrodynamic calculations, however, show that this is not a realistic assumption, because some material spills over past the truncation radius. This material may still contribute to the disk observables \citep{panoglou}. 

To compare the model with observations, we additionally need the value of interstellar reddening $E(B-V)$ and two geometrical parameters: the inclination angle $i$ (0{\degree} for pole-on orientations) and the distance $d$. 

\begin{table*}[t]
 \caption[]{\label{adopted_par}Adopted model parameters.}
\centering
\begin{tabular}{lcccc}
 \hline \hline
  Star & Spectral type\tablefootmark{a,b} & Mass\tablefootmark{c} [M$_\odot$] & Distance\tablefootmark{d} [pc] & Inclination [{\degree}] \\
\hline
$\beta$ CMi    & B8\,Ve       & 3.5\tablefootmark{e}  & 49.6 & 43\tablefootmark{e} \\ 
$\eta$ Tau    & B7\,IIIe       & 3.4  & 123.6 & 41\tablefootmark{f} \\ 
EW Lac        & B3\,IVe-shell  & 6.1  & 252 & 80\tablefootmark{b,g} \\ 
$\psi$ Per    & B5\,IIIe-shell & 4.4  & 178.9 & 75\tablefootmark{h} \\ 
$\gamma$ Cas  & B0.5\,IVe      & 13.2 & 168.4 & 42\tablefootmark{i}\\ 
$\beta$ Mon A & B4\,Ve-shell   & 5.1  & 208 & 70\tablefootmark{b,g} \\ 
\hline
\end{tabular}
\tablefoot{}
\tablefoottext{a}{\citet{slettebak};
\tablefoottext{b}{\citet{rivinius2006};}
\tablefoottext{c}{\citet{harmanec};}
\tablefoottext{d}{\citet{hip};}
\tablefoottext{e}{\citet{klement};}
\tablefoottext{f}{\citet{tycner2005};}
\tablefoottext{g}{This work;}
\tablefoottext{h}{\citet{delaa};}
\tablefoottext{i}{\citet{stee2012}.}
}
\end{table*}

\begin{table*}[t]
 \caption[]{\label{bestfit_par}Derived model parameters.}
\centering
\begin{tabular}{lcccccccc}
 \hline \hline
  Star & $R_\text{p}$ [R$_\odot$] &  $R_\text{e}$ [R$_\odot$] & $L$ [L$_\odot$] & $\log{\rho_0}$ [g\,cm$^{-3}$] & $n$ & $R_\text{out}$ [R$_\text{e}$]& $R_\text{c}$  [R$_\text{e}$]& $E(B-V)$ \\
\hline 
$\beta$ CMi   & $2.8 \pm 0.2 $ & $4.20$   & $185 \pm 5$   & $-11.78^{+0.18}_{-0.30}$ & $2.9 \pm 0.1$ & $40^{+10}_{-5}$  & 400 & $0.01^{+0.02}_{-0.01}$\\
$\eta$ Tau    & $8.0 \pm 0.5$   & $10.56$ & $1750 \pm 50$  & $-12.23^{+0.12}_{-0.16}$ & $3.0 \pm 0.4$ & $40^{+10}_{-5}$     & 140       & $0.06\pm 0.02$\\
EW Lac        & $4.5 \pm 0.5$ & $5.94$  & $3500 \pm 100$  & $-11.60^{+0.12}_{-0.18}$ & $2.2 \pm 0.1$ & $100^{+30}_{-20}$    & 360       & $0.11^{+0.04}_{-0.03}$ \\
$\psi$ Per    & $5.5 \pm 0.5$ & $7.26$  & $2100 \pm 100$  & $-11.48^{+0.10}_{-0.12}$ & $2.5 \pm 0.1$ & $100^{+5}_{-15} $ & 250 & $0.08^{+0.01}_{-0.02}$ \\
$\gamma$ Cas  & $7.0 \pm 0.5$   & $9.24$  & $19000 \pm 500$ & $-10.48^{+0.10}_{-0.12}$ & $3.3 \pm 0.1$ & $35 \pm 5$ & 560 & $0.04^{+0.03}_{-0.02}$ \\
$\beta$ Mon A & $4.0 \pm 0.5$   & $5.28$  & $4500 \pm 100$ & $-10.60^{+0.19}_{-0.18}$ & $3.0^{+0.2}_{-0.1}$ & $110 \pm 40$  & 380       & 0.11$\pm$ 0.04\\
\hline
\end{tabular}
\tablefoot{The best-fit model parameters of the central star ($R_\text{p}$, $L$), the disk ($\log{\rho_0}$, $n$, $R_\text{out}$), and the interstellar medium ($E(B-V)$). The calculated values of $R_\text{e}$ (for given $W$), and $R_\text{c}$ (Eq.~\ref{krticka}) are also listed.
}
\end{table*}

\subsection{Modeling procedure}

To calculate the synthetic observables, we employ the code {\ttfamily HDUST} \citep{hdust1, hdust2}. All the objects are modeled as a combination of a fast-rotating central star surrounded by a VDD. The code uses the Monte Carlo method to solve the radiative transfer, radiative equilibrium, and statistical equilibrium in 3D geometry for the given density and velocity distributions for pure hydrogen gas. For details on the hydrodynamics of VDDs, we refer to \citet{carciofi2011} and \citet{krticka}.

The modeling procedure is similar to the one tested by \citet{klement} when using a multi-technique and multi-wavelength dataset. In the present work we used only the SED flux measurements to constrain the models. As this limits the constraining ability of the modeling, we keep several parameters fixed, specifically the stellar mass, rotation rate, inclination and distance, and we focus instead on determining the parameters that are best constrained by the SED structure, namely the interstellar reddening, the stellar radius and luminosity, the disk parameters ($\rho_0$, $n$) and the outer disk radius. 

The multi-epoch data were combined into a single dataset, as the variability of the studied stars was found to be reasonably low. Where signs of variability are present, the combined data represent average properties of the disks in the last decades (for details see the results for individual targets in Sect.~\ref{results}).

We started by searching the literature for spectral type classifications of the programme stars. The spectral types were used to fix the masses of the central stars as given by \citet{harmanec} for the fundamental parameters of main sequence B-type binaries. The rotation rate was kept at \text{the typical value} of $W$ = 0.8 \citep{review}. This value corresponds to a $R_\text{e}/R_\text{p}$ ratio of 1.32. The gravity darkening exponent was correspondingly set to 0.1728, as given by \citet{lara}. The distances to the stars were taken from the Hipparcos measurements \citep{hip}, and the inclinations from interferometric studies, when available. For the shell stars, we assumed $i > 70${\degree}.The adopted parameters are listed in Table~\ref{adopted_par}.

Interstellar reddenning presents an additional model parameter $E(B-V)$, which needs to be determined in order to be able to compare the observations to the model. We used the 2200\,{\AA} absorption bump (caused by the interstellar medium) to determine the $E(B-V)$ of the programme stars \citep[we refer to][for discussion of the origin of the bump]{zagury}. During the procedure the observed UV spectra are dereddened using the reddening curve of \citet{1999PASP..111...63F} with $E(B-V)$ as a free parameter, until the 2200\,{\AA} bump disappears and the spectrum becomes flat in that region. For the extinction $R_V$ we used the standard value of 3.1 \citep{savage}. Although values other than 3.1 may be observed in nebular and star-forming regions, for nearby field stars such deviations are less likely. The uncertainties of the determined values of the $E(B-V)$ parameter were inferred by a Markov Chain Monte Carlo (MCMC) method provided by the emcee \textsc{Python} module\footnote{available online at \url{http://dan.iel.fm/emcee} under the MIT License} \citep{2013PASP..125..306F}.

We proceed to the remaining physical parameters of the central star, namely the polar radius $R_\text{p}$ and luminosity $L$. As discussed in \citet{klement}, for non-shell stars or shell stars with tenuous disks, the disk has very little influence on the observed UV spectrum. In such cases and within a reasonable parameter range, changing $R_\text{p}$ (which in effect means changing the effective temperature as well) influences the slope of the model UV spectrum only, while changing $L$ influences mainly its level. This allows us to constrain these two parameters independently by computing a grid of purely photospheric, fast-rotating models \citep[similar to][]{klement}. Furthermore, for tenuous disks, the visual and near-IR parts of the SED are also devoid of disk influence. The wavelength interval used to determine $R_\text{p}$ and $L$ was therefore selected on a case-by-case basis (we refer to Sect.~\ref{results} for details on individual targets).

The best-fit $R_\text{p}$ and $L$ parameter combination was determined via $\chi^2$ minimization, where for the best-fit model $\chi^2= \chi^2_\text{min}$. Since we were dealing with SED modeling, where both the wavelengths and fluxes span many orders of magnitude, we used the logarithmic version of $\chi^2$:
\begin{equation}
\chi^2_\mathrm{mod} = \sum_{i=1}^{N} \left(\frac{\log{F_\text{obs,i}} - \log{F_\text{mod,i}}}{F_\text{err,i}/{F_\text{obs,i}}}\right)^2,
\end{equation}
where $F_\text{obs}$ is the observed flux, $F_\text{mod}$ is the model flux, $F_\text{err}$ is the observed flux error, and $N$ is the total number of fitted data points. The uncertainties of the best-fit parameters were estimated from the $\Delta\chi^2_{0.683}=\chi^2_{\mathrm{mod}} - \chi^2_{\mathrm{min}}$ contour corresponding to the $68.3\%$ confidence level (1-$\sigma$), where $\Delta\chi^2_{0.683}$ is a function of the number of degrees of freedom of the fit \citep[we refer to, e.g., Chapter 11 of][]{bevington}.  In some cases the parameter uncertainties turned out to be lower than the step size adopted for the parameter grid. Since the quality of the model fit was already sufficient for the aims of this work, we did not put in the computational effort to further fine-tune the model parameters and thus lower the uncertainty estimates, and we instead conservatively selected the corresponding step size as the parameter uncertainty. 

The disk itself is non-isothermal with a temperature structure computed by {\ttfamily HDUST}, but its density structure is not solved self-consistently with the temperature solution and remains the same throughout the simulation. For such a simplified VDD model we need only three free parameters to fully describe it: the base density $\rho_0$, the density exponent $n$, and the disk outer radius $R_\text{out}$. We set an upper limit to the $R_\text{out}$ parameter, so that it cannot exceed the value of the critical radius $R_\text{c}$ estimated by means of Eq.~(\ref{krticka}), in which the sound speed $c_\text{s}$ was computed using the mass-averaged temperature of the disk.

In our model, we assume a single value of $n$ throughout the whole disk and by doing so we are neglecting local non-isothermal density effects in the disk. VDD theory predicts a temperature minimum to occur in the inner parts of the disk due to its high opacity to the stellar radiation. However, since we are interested in the overall density structure of the VDD, of which the inner non-isothermal part is only a small fraction, these effects are not important for our analysis. 

The modeling procedure was executed as follows. First, we fixed $R_\text{out}$ to an intermediate value of 50 stellar radii, and a grid of models with $\rho_0$ and $n$ in the range 1$\times10^{-13}$ -- 1$\times10^{-10}$\,g\,cm$^{-3}$ and 1.5 -- 5.0, respectively, was computed. To determine the parameters $\rho_0$ and $n$, we fit the model grid to the part of the SED where the disk contribution is significant, but not to the long wavelength radio data, that are sensitive to the disk size. While dense disks start to contribute to the SED already at visual wavelengths, excess radiation due to a tenuous disk might become observable only at mid-IR wavelengths. Similarly, the truncation effects in very dense disks will influence the SED at shorter wavelengths as compared to less dense disks. Therefore, the SED region used to constrain $\rho_0$ and $n$ will be different from star to star and will shift towards shorter wavelengths with increasing density of the disk (for details on individual targets see Sect.~\ref{results}). The uncertainties of the best-fit values of the parameters $\rho_0$ and $n$ were determined in the same way as for the central star parameters $R_\text{p}$ and $L$, described above.

The modeling procedure gets more complicated for shell stars with dense disks. In these cases, the disk reduces the UV flux (and sometimes also portions of the visual and near-IR flux) emitted by the central star and the observed UV spectrum cannot, therefore, be used to constrain $R_\text{p}$ and $L$ in a straightforward way. We adopted an iterative modeling procedure for such stars: A first look estimate for $R_\text{p}$ and $L$ was done using the photospheric models only, after which the disk was included in the model to get the first estimates for the disk parameters $\rho_0$ and $n$ from the IR fluxes. With the values obtained in this way, the magnitude of the influence of the disk on the UV spectrum could be estimated and taken into account for the next iteration of photospheric parameters. After a few such iterations, a satisfactory fit to both the UV and IR parts of the SED was obtained.

After determining the best-fit values of $\rho_0$ and $n$, we can proceed to the determination of the $R_\text{out}$ parameter. As mentioned above, the truncation effects will become detectable at progressively shorter wavelengths with increasing density of the disk. However, with disk sizes $\gtrsim20$\,$R_\text{e}$, the influence does not fall shortwards of sub-mm/mm wavelengths. Therefore, the mm and cm fluxes are suitable for constraining the sizes of dense disks, while cm fluxes only constrain the disk size of the tenuous ones (e.g., $\beta$~CMi). In the latter case, the sub-mm/mm region was included when searching for the best-fit values of $\rho_0$ and $n$. Measurements corresponding to upper limits were used for visual checking, but not for the modeling procedure itself. The uncertainty of the $R_\text{out}$ parameter was obtained in the same way as for the previously mentioned model parameters.


\section{Results}\label{results}

In this section we first review the results for $\beta$~CMi from \citet{klement}. Then we present the results for stars for which we have new JVLA measurements ($\eta$~Tau, EW~Lac, $\psi$~Per, $\gamma$~Cas), followed by the results for $\beta$~Mon~A, which has new sub-mm data from APEX/LABOCA. We also give a brief overview of what can be learned from the available visual spectroscopy and polarimetry for each target. The derived model parameters for all stars are listed in Table~\ref{bestfit_par}. In Table~\ref{bestfit_par} we also include the values of $R_\text{c}$ calculated using Eq.~(\ref{krticka}). 

      \begin{figure*}[t!]
   \centering
   \includegraphics[width=\hsize]{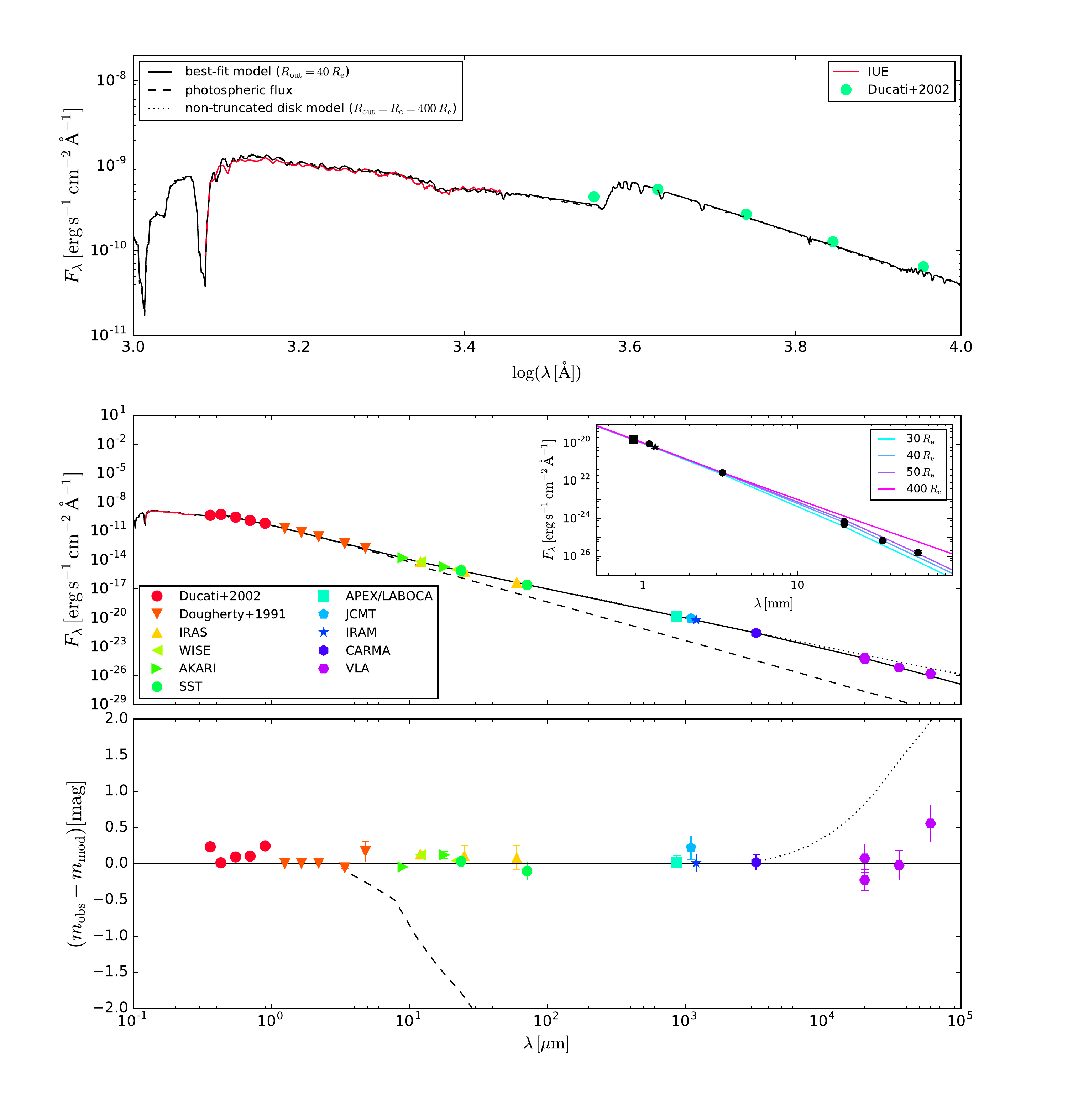}
      \caption{$\beta$~CMi. The best-fit model is plotted as the black solid line, the photospheric contribution as the dashed black line, and a non-truncated disk model ($R_\text{out} = R_\text{c}$) as the black dotted line. \textit{Upper:} The SED in the 1000--10000\,$\AA$ interval. \textit{Middle:} The SED from 0.1\,$\mu$m to 10\,cm. The inset shows model predictions for different disk sizes. The observed fluxes are plotted with error bars, which are smaller than the symbol size in most cases. \textit{Lower:} The residuals structure of the best-fit model with $R_\text{out} = 40\,R_\text{e}$. The photospheric and non-truncated models are also plotted in relation to the best-fit model, highlighting where these models deviate from the best-fit one.
              }
         \label{bcmi}
   \end{figure*}

   \begin{figure*}[t!]
   \centering
   \includegraphics[width=\hsize]{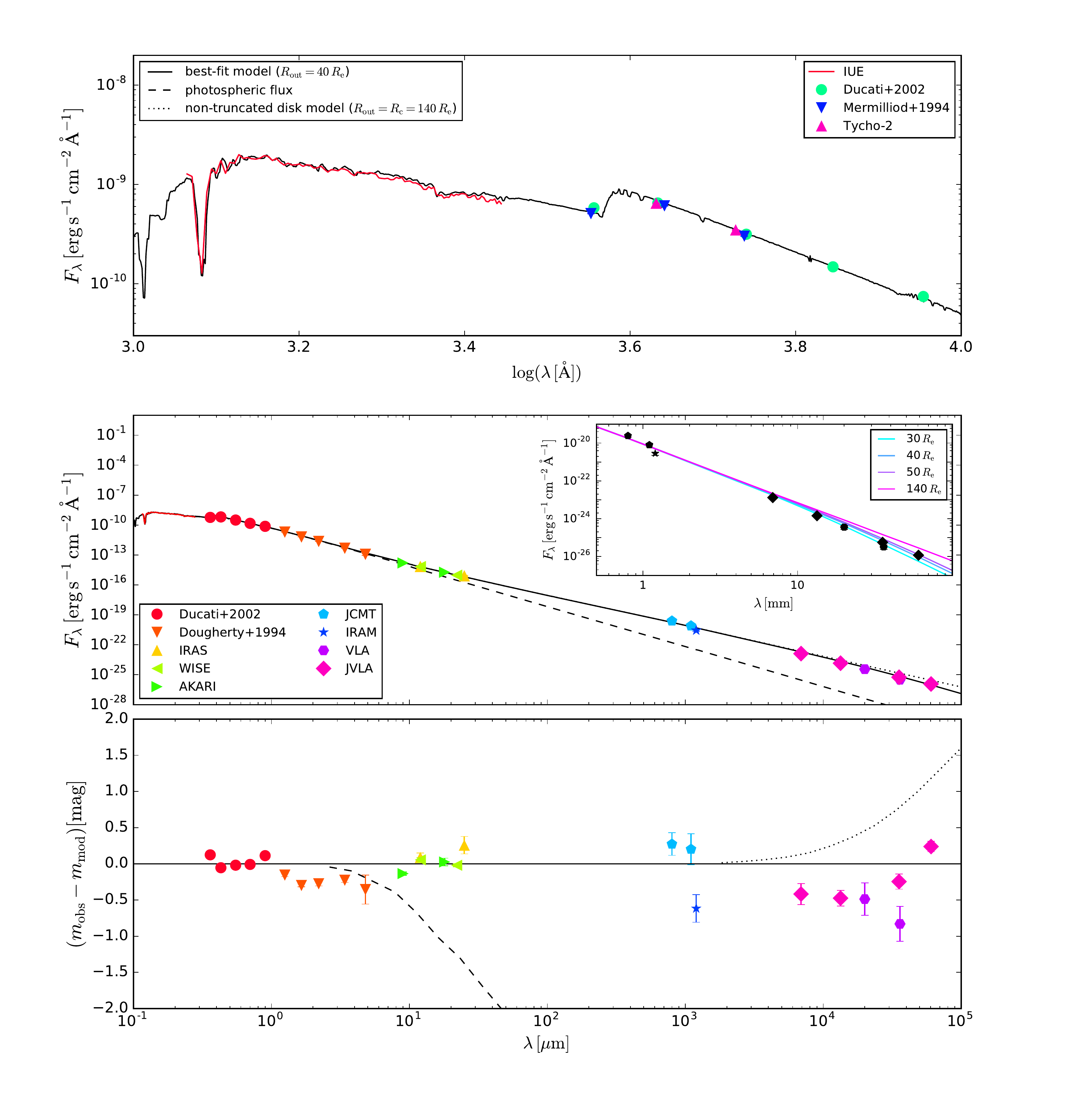}
      \caption{As in Fig.~1, but for $\eta$ Tau.
}
         \label{etau}
   \end{figure*}

\subsection{$\beta$~CMi (HD 58715; HR 2845)}

The star $ \beta$~CMi was studied in extensive detail by \citet{klement}. As in the present study, the {\ttfamily HDUST} code was used to model the SED from the UV to the radio, but additionally a large data set of spectroscopic, polarimetric, and interferometric observations was used to further constrain the model. Here we present a slightly adjusted model of the SED produced using an updated version of the {\ttfamily HDUST} code, which fixed some numerical problems that were slightly affecting the correct computation of synthetic observables. More details on the fixes of the code and the changes regarding the results of modeling of the observables besides the SED are given in \citet{klement_aspc}.

We recomputed the model and allowed the $\rho_0$, $n$ and $R_\text{out}$ parameters to vary in a small interval around the values originally determined by \citet{klement}, while keeping the central star parameters the same as in the previous study. The best-fit value of base density $\rho_0$ remains the same, but a significant change to the model concerns the density exponent $n$: the SED up to mm wavelengths is now reproduced exceptionally well when using a value of $n = 2.9$. We recall that in the original study, the very inner parts of the disk were compatible with a steeper ($n = 3.5$) density fall-off, while in the remainder of the disk it followed a shallower profile ($n = 3.0$).

Regarding the outer parts of the disk, the original conclusion was that the disk is truncated at a distance of $35^{+10}_{-5}$\,$R_\text{e}$, as revealed by the observed radio flux at 2\,cm. Here we include three additional VLA measurements at cm wavelengths that are available from the literature. The revised model and additional cm data led to a revised best-fit disk size of $40^{+10}_{-5}$\,$R_\text{e}$. The best-fit model is plotted in Fig.~1 as a solid black line, together with the purely photospheric flux (dashed black line) and a non-truncated disk model (dotted black line). The model reproduces the full SED almost perfectly, with the exception of the data point at 6\,cm, which lies slightly above our best-fit model.

\subsection{$\eta$ Tau (Alcyone; HD 23630; HR 1165)}


This star from the Pleiades cluster has been mostly classified in the literature as a B7IIIe star \citep{slettebak,lesh,hoffleit}. The inclination is apparently of an intermediate value. In previous interferometric studies it was quoted to be higher than 18{\degree} \citep{quir} and equal to 41{\degree} \citep{tycner2005}. We adopt the latter value for our model.

No companion was revealed in adaptive optics observations \citep{roberts2007}. A total of 47 spectra suitable for RV measurement were found in the archives. The RV was measured in the \ion{Mg}{ii}\,4481 line. The mean RV is $6.9\pm2.1$\,km/s. {\sc Heros} measurements around MJD~52000 are on average approximately 4\,km/s lower than ESPaDOnS measurements at MJD~55000, but given the different instrument parameters, the values are well compatible and within 3-$\sigma$ of one another. However, we note that this by no means excludes the possible presence of a binary companion. The IR Ca triplet emission, indicative of binarity, is not present in the spectra.  

The spectroscopic data show that the H$\alpha$ emission was at a constant level of $I/I_\mathrm{c}=2.2$, and $W_\lambda=-3.5$\,{\AA} until approximately 2005, and since then steadily decreased by a small but detectable amount to $I/I_\mathrm{c}=2.0$, and $W_\lambda=-2.5$\,\AA. The spectropolarimetric measurements available from the HPOL database show no change of polarization degree or angle between 1992 and 1999.

From the comparison of IRAS and AKARI/WISE measurements, which are separated in time by around two decades, we also see that the variability of the disk has been relatively low. This is a common feature of late-type Be stars. The comparison of the observed mm fluxes from JCMT and IRAM shows signs of slight variability at mm wavelengths on the time scale of years. The archival VLA and the new JVLA measurements are very much consistent with one another, with the exception of the observed fluxes at 3.6\,cm, which however still lie within 2-$\sigma$ of one another.

  \begin{figure*}[t!]
   \centering
   \includegraphics[width=\hsize]{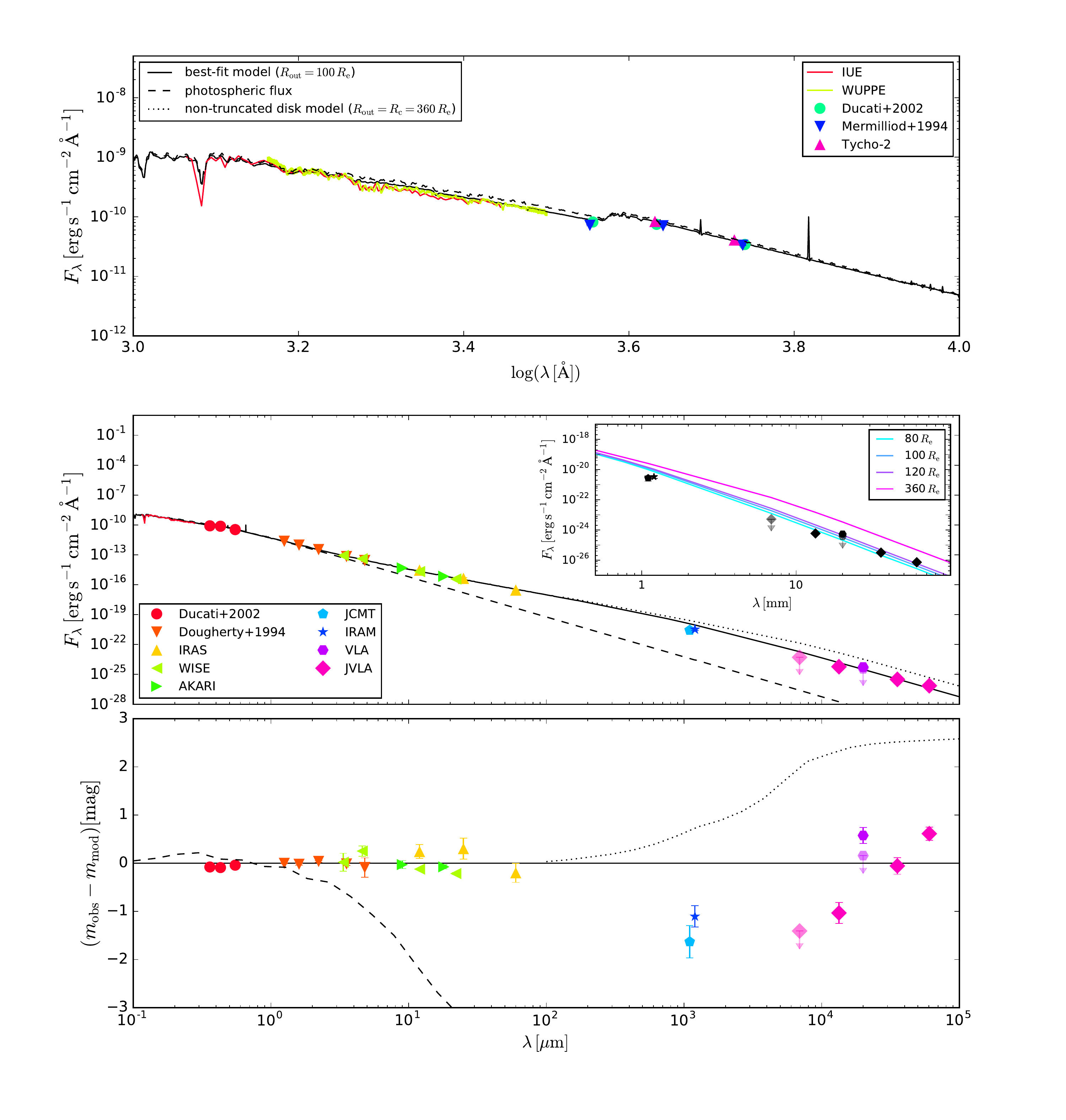}
      \caption{As in Fig.~1, but for EW Lac.
              }
         \label{ewla}
   \end{figure*}

  \begin{figure*}[t!]
   \centering
   \includegraphics[width=\hsize]{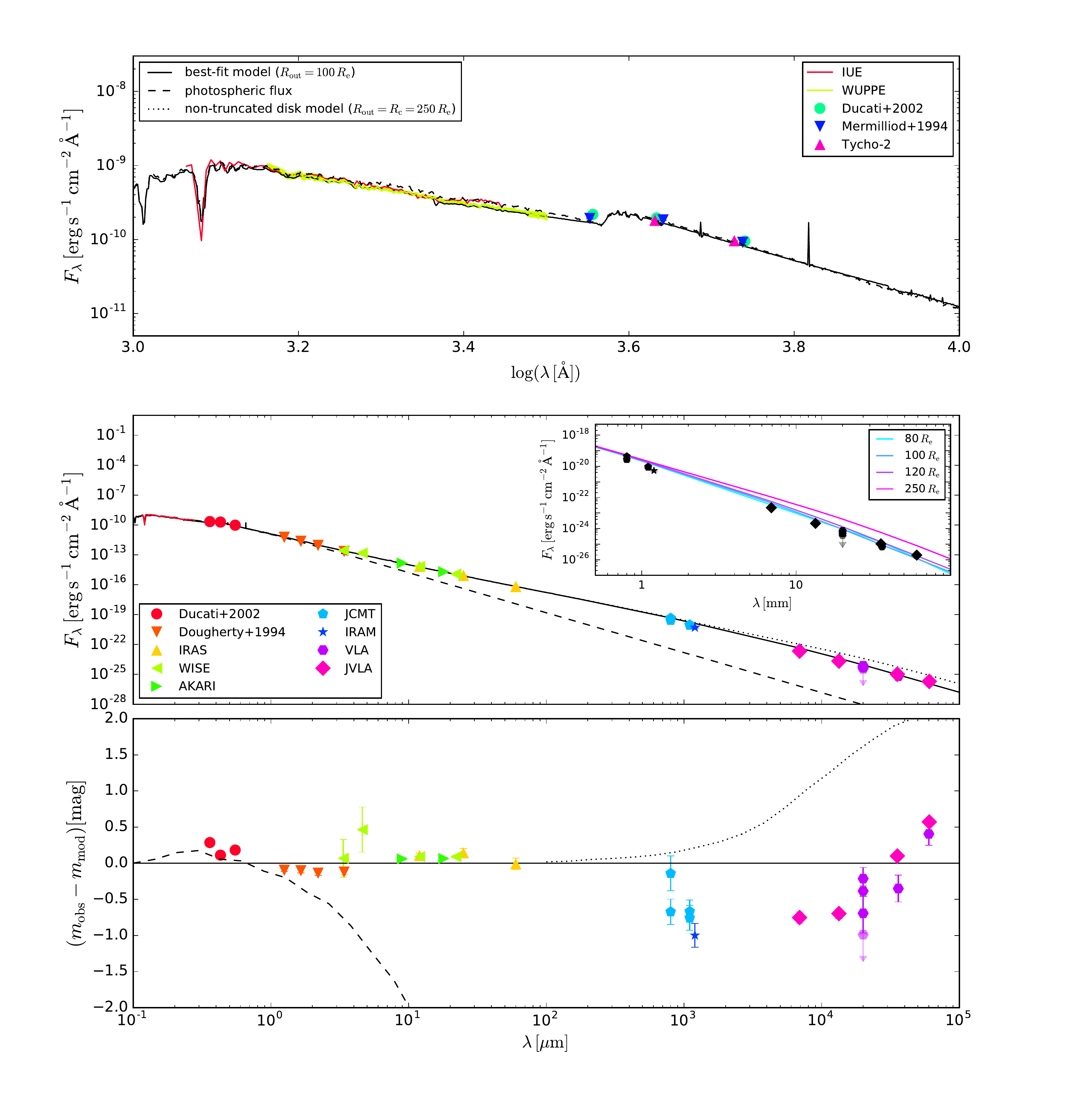}
      \caption{As in Fig.~1, but for $\psi$ Per.
              }
         \label{pper}
   \end{figure*}

The results of the SED modeling of $\eta$~Tau are plotted in Fig.~\ref{etau}. The non-shell nature of $\eta$ Tau allows us to use the UV spectrum to determine the central star parameters $R_\text{p}$ and $L$ without any complications. Moreover, the disk of $\eta$ Tau is apparently very tenuous and starts to contribute to the observed SED only at around 5\,$\mu$m. We thus use both the UV spectrum and the visual photometry to constrain $R_\text{p}$ and $L$. Changing the disk size influences the cm fluxes only and has a negligible influence on the mm fluxes (as expected for tenuous disks). Therefore, the SED interval from 5\,$\mu$m to mm was used to find the best-fit values of the disk parameters $\rho_0$ and $n$. 

The observed radio fluxes are lower than expected from a non-truncated VDD model. The disk size best reproducing the cm data is $R_\text{out} = 40^{+10}_{-5}$\,$R_\text{e}$. We interpret this as a clear sign of truncation of the outer disk. However, a simply truncated VDD does not offer a good fit either, as the observed SED slope in the radio is flatter than the model predicts. This is most easily seen in the residuals of the best-fit model with respect to the radio measurements (bottom panel of Fig.~\ref{etau}). 

\subsection{EW Lac (HD 21750; HR 8731)}

EW Lac was classified as a B3IVe-shell star \citep{slettebak, rivinius2006}, and a B4IIIep star \citep{lesh, hoffleit}. No previous interferometric determination of the disk inclination exists in the literature. However, the shell nature of the star restricts the inclination angles to $\geq$70{\degree}. For our study we adopt the value of $i$ = 80{\degree}, which was found to reproduce well the shapes of the Balmer and Paschen jumps.

EW Lac has shown remarkable variations in its shell line appearance in the past, most notably from 1976 to 1986, and a similar episode lasted from 2007 until 2012 \citep{mon2013}. The variations are best explained in terms of V/R variability, that is, changes in the peak height ratio of the violet and red side of the emission. Small V/R differences were present in almost all spectra at our disposal, taken from 1998 to 2014. However, beginning in the 2007 to 2014 period, that is, for the last V/R cycle, the amplitudes were greatly enhanced.

There are 13 suitable spectra to measure RVs, from which the presence of a companion could be neither confirmed, nor excluded. Values measured in the \ion{Fe}{ii}\,5169 line obtained from {\sc Heros} spectra, from MJD~51000 to 52500, are at approximately $-22$\,km/s, while ESPaDOnS values, obtained around MJD~55400, are at $-28$\,km/s. Although this change is very likely real, it is probably due to a change in the global density oscillation of the disk rather than binarity. Also, changes in the H$\alpha$ equivalent width are due to the variable strength of the central absorption core, rather than changes in the strength of the emitting peaks. 

Like for $\eta$\,Tau, the HPOL data, taken from 1989 to 1995, do not show any change in polarization degree or position angle; neither is the infrared Ca triplet observed in emission in our spectra. 

Comparison of the IRAS and AKARI/WISE measurements reveals a variation in the IR excess emission that is somewhat stronger than in the other targets in our sample. The mm fluxes observed by JCMT and IRAM are consistent within the respective error bars. The comparison of the VLA and JVLA measurements also reveals a slight discrepancy in that one of the VLA measurements lies a few $\sigma$ above the JVLA fluxes. The two VLA observations at 2\,cm are slightly inconsistent, as the upper limit measurement lies below the detection. This suggests a possible mild variability of cm fluxes on the time scale of a year. 

The results of the SED modeling of EW~Lac are plotted in Fig.~\ref{ewla}. Due to the shell nature of the star and the high value of the disk density, the observed UV spectrum of EW Lac is slightly dimmed by the surrounding disk. The disk excess starts to be significant at a wavelength of around 1\,$\mu$m, allowing for the inclusion of the visual part of the SED for determining the central star parameters. Since the disk is dense, the IR part of the SED was used to determine $\rho_O$ and $n$, while the mm and radio part, sensitive to varying the disk size, was used to determine $R_\text{out}$. The resulting SED fit up to far-IR is satisfactory.

The radio observations, including the mm measurements, are inconsistent with a non-truncated disk. The resulting disk size for the case of simple truncation is $100^{+30}_{-20}$\,$R_\text{e}$. Moreover, the new JVLA data set reveals a similar result as given above for $\eta$~Tau -- the observed slope of the SED in the radio is not as steep as would be expected for a simply truncated disk.

\subsection{$\psi$ Per (HD 22192; HR 1087)}

The star $ \psi$\,Per has been classified as a B5Ve-shell star \citep{lesh, hoffleit}, a B4IVe star (Underhill 1979) and also as a B5IIIe star \citep{slettebak, rivinius2006}. The inclination angle is $> 62${\degree} according to \citet{quir} and $75${\degree} $\pm$ $8 ${\degree} according to \citet{delaa}; for this study we adopt the latter value.

No binary companion is evident (Mason et al. 1997; Delaa et al.  2011). Yet, for this star, the infrared Ca triplet is observed in clear emission in all available spectra covering that region. Like for EW\,Lac, the shell nature permits the measurement of the RV with a precision of a few dozen m/s. While the RV is varying by approximately 1500\,m/s peak to peak, there are not enough spectra to say whether this is due to a companion, that is, clearly periodic, or due to small changes in the disk, in which case changes could be merely cyclic or secular.

For $\psi$\,Per, the HPOL data show a clear increase of polarization degree from 0.4\% to 0.8\% from 1992 to 1994, which then remains at 0.8\% until at least 2000. All available spectra were taken in 2000 to 2014, that is, after the polarization changed, and show a strong and stable disk, with $I/I_\mathrm{c}=6$ and $W_\lambda=-41$\,\AA. The V/R ratio of the emission lines is unity; apparent deviations observed, for example, in \ion{Fe}{ii}\,5169, are due to blends.

No variability in the IR excess is revealed by comparing the IRAS and AKARI/WISE data. The mm measurements from JCMT and IRAM are mostly consistent, although one of the JCMT detections lies slightly above the others. The old VLA and the new JVLA data are in good agreement, although the upper limit of \citet{apparao} at 2\,cm lies below the subsequent detections. It is unclear whether this is caused by radio variability or by the low signal-to-noise ratio of the observations.

The results of the SED modeling of $\psi$ Per are plotted in Fig.~\ref{pper}. Although $\psi$ Per is a shell star, the disk base density is only of an intermediate value and does not significantly influence the UV and visual part of the SED. Therefore the SED interval up to 1 micron was used to constrain the central star $R_\text{p}$ and $L$. The disk contribution to the observed SED starts to be significant at near-IR wavelengths. Although the value of $\rho_0$ is only intermediate, the fact that $n$ is very low makes the disk dense in its outer parts. Correspondingly, the model SED was found to be sensitive to changing the disk size already at mm wavelengths, therefore it was only the IR portion of the spectrum that was used to constrain the disk density parameters. The resulting fit from UV to far-IR wavelengths is exceptionally good.

The observed radio SED slope is once again inconsistent with both a non-truncated and a simply truncated disk. The resulting disk size best reproducing the whole radio data set is $100^{+5}_{-15}$\,$R_\text{e}$. The result again suggests that the disk density fall-off regime changes to a steeper one in the outer parts, but the disk is not simply cut-off, as is assumed in our model.

\begin{figure}[t!]
   \centering
   \includegraphics[width=\hsize]{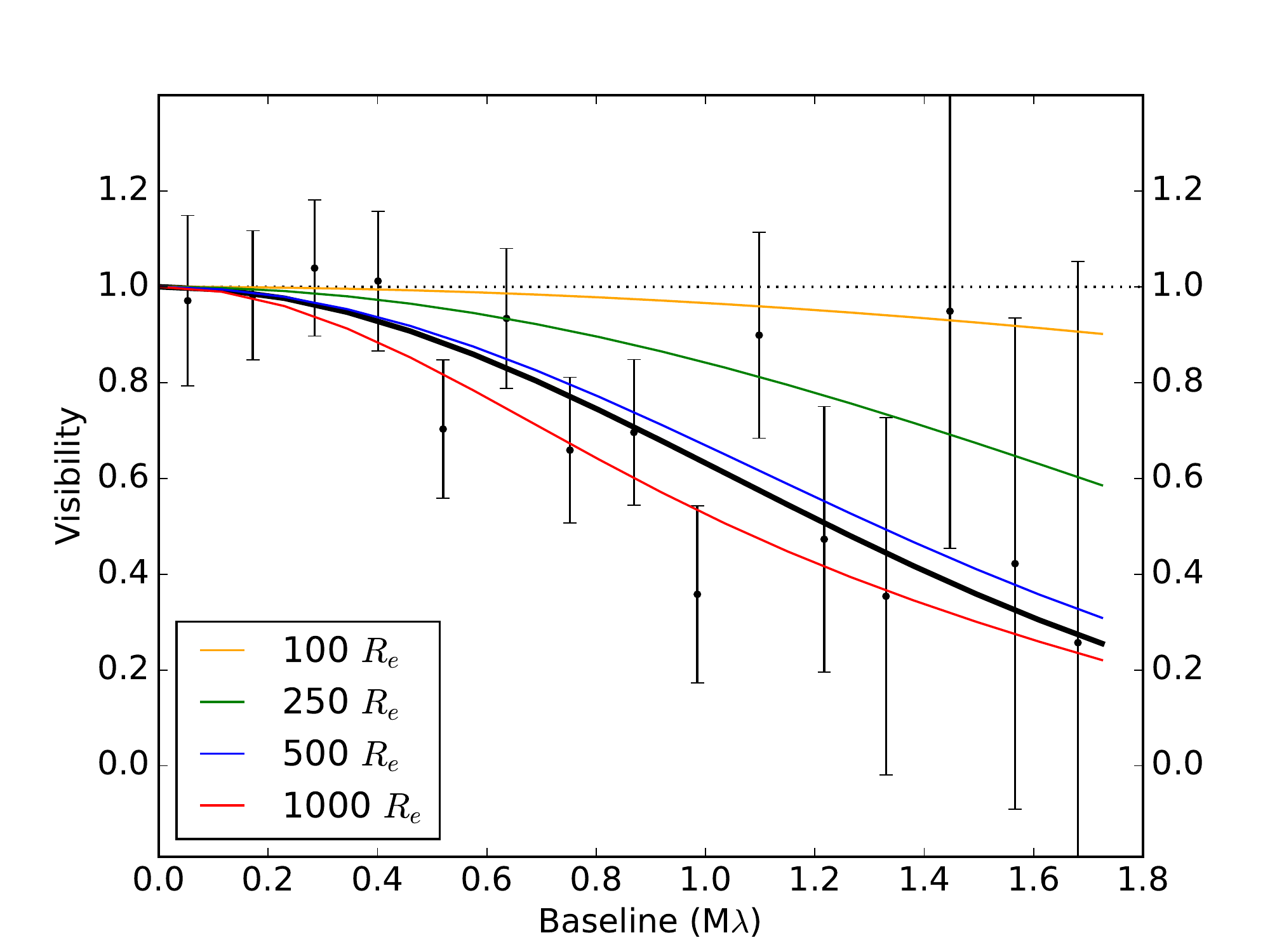}
      \caption{$\psi$~Per: The azimuthally averaged visibility data of \citet[][black points with error bars]{dougherty_nature} and a Gaussian fit to them (thick black line, FWHM of 74\,mas corresponding to 518\,$R_\text{e}$) over-plotted with the visibility curves derived from our azimuthally averaged models with several different disk sizes indicated in the legend. The dotted line shows the visibility curve of a point source (visibility equal to unity with increasing baseline length).
              }
         \label{dt1992}
   \end{figure}

To this date, $\psi$\,Per is the only star that has been angularly resolved in radio ($\lambda$ = 2\,cm) along the semi-major axis of its disk \citep{dougherty_nature}. In that study, the Gaussian fit to the azimuthally averaged visibility data indicated a best-fit disk size (FWHM) of 74 mas, corresponding to 392\,$R_\text{e}$. In Fig.~\ref{dt1992} we plot the original visibility curve of \citet{dougherty_nature}, along with the curves derived from our models with different disk sizes (azimuthally averaged). Since the absolute flux density values for each of our models differ, we scale the curves so that at zero baseline the visibility is equal to unity. The visibility curve of our best-fit model (R$_\text{out} = 100$\,$R_\text{e}$) shows that such a disk would not have been resolved by the observations of \citet{dougherty_nature}, although the model flux density at 2\,cm (1190\,$\mu$Jy) is relatively close to the measured value ($804 \pm 60$\,$\mu$Jy). The model with the disk size of $500$\,$R_\text{e}$ shows almost perfect agreement with the shape of the observed visibility curve. However, the flux density of this model (9.31\,mJy) overestimates the measured value by more than a factor of 10. As for the apparent disk sizes of our models, 2D Gaussian fits give FWHMs of $29.5\times 20.1$\,mas and $85.2\times 47.8$\,mas for the best-fit and the $500$\,$R_\text{e}$ disk models, respectively. The disagreement between our model and the angular size and fluxes of \citet{dougherty_nature} cannot be solved by the simple models that we employ, and will be the subject of future research.

\subsection{$\gamma$ Cas (HD 5394; HR 264)}

The star $ \gamma$ Cas was one of the first two Be stars to be discovered \citep{secchi}. It was identified as an X-ray source and there is an ongoing debate as to the origin of the puzzling X-ray emission properties \citep[e.g., ][]{motch}. $\gamma$ Cas is a known single-line spectroscopic binary. The set of orbital parameters was recently revised by \citet{nemravova2012}, who concluded that the orbit is circular with an RV semi-amplitude of $4.30 \pm 0.09$\,km/s and a period of $203.52 \pm 0.08$ days. $\gamma$ Cas has a very complicated observational history, as it has appeared as a shell star, briefly as a normal B star and since a few decades ago as a Be star seen under intermediate inclination \citep[see][for details]{harmanec2002}. 

$\gamma$\,Cas has been classified in the literature as a B0.5IVe star \citep{slettebak}. For the inclination, we adopt the value of 42{\degree}, found in a recent interferometric study by \citet{stee2012}. 

In agreement with the suspicion that the infrared Ca triplet may be linked to binarity \citep{2011JPhCS.328a2026K}, $\gamma$\,Cas does indeed show emission in these spectral lines. The strength is somewhat lower than that of the Paschen lines that the Ca triplet lines are blended with, but clearly discernible. The shape of the H$\alpha$ line also seems to indicate binarity, since it often shows no clear peaks or symmetry, which are signs of disturbances in the disk.

   \begin{figure*}[t!]
   \centering
   \includegraphics[width=\hsize]{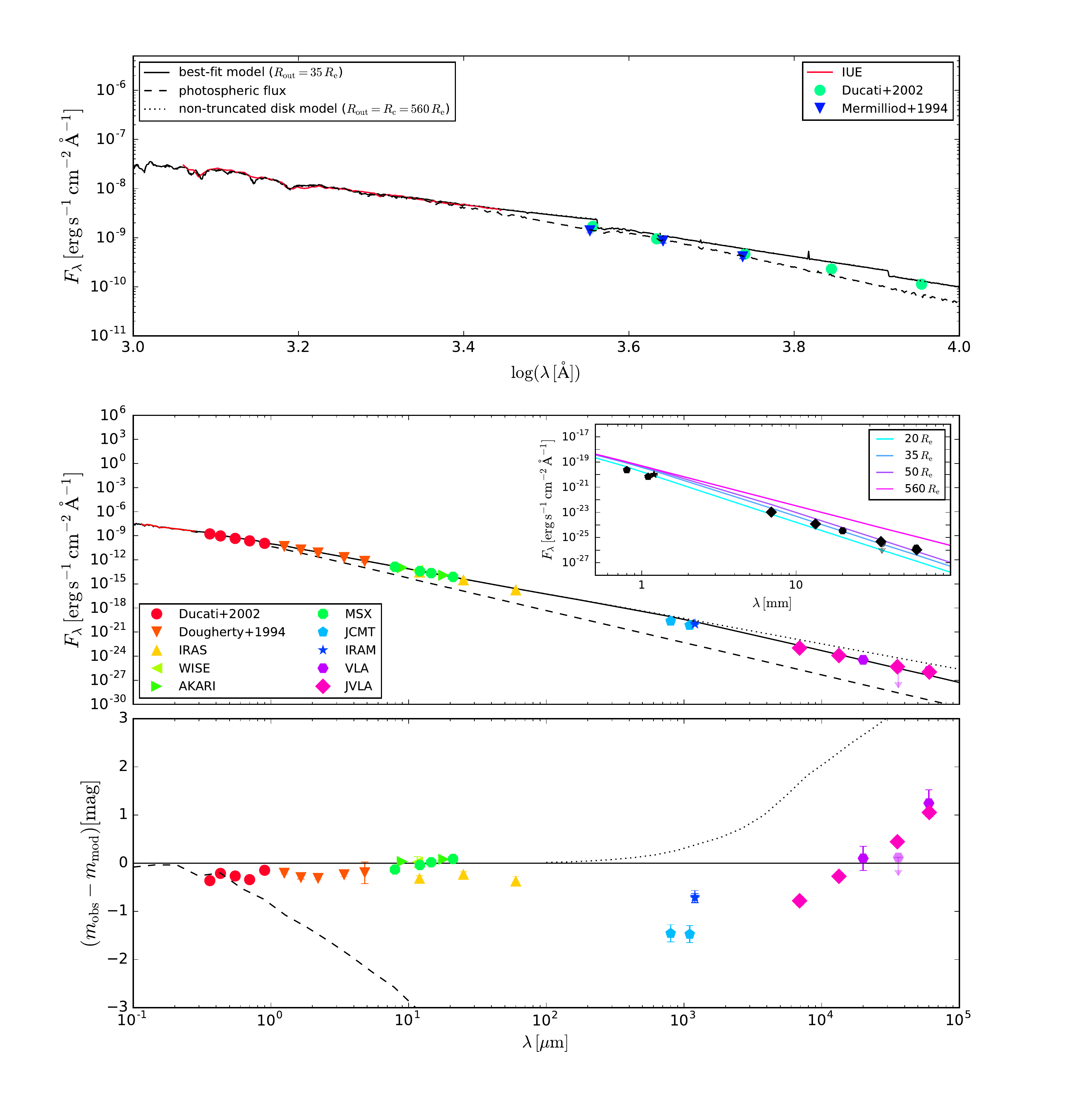}
      \caption{As in Fig.~1, but for $\gamma$ Cas.
              }
         \label{gcas}
   \end{figure*}

Judging from the H$\beta$ profiles, which behave closer to the expectations for a classical Be star than the H$\alpha$ profiles, $\gamma$\,Cas underwent V/R variability before 2000, which weakened between 2000 and 2005, and is since
absent. The available HPOL data, spanning 1991 to 2000, show a constant polarization. Since $\gamma$\,Cas is the brightest Be star in the Northern hemisphere, there is a great wealth of spectroscopic data available from the archives. Here spectra taken between 1996 and 2014 are considered. The value for $I/I_\mathrm{c}$ of H$\alpha$ varied, but remained in the range between approximately 4 and 5. In other words, while the disk of $\gamma$\,Cas is not stable in the same sense as the one of, for instance, $\psi$\,Per, it is clearly not currently in a stage of secular build-up or decay, but fluctuating around a certain state.

The IR measurements do however reveal signs of variability, as the IRAS fluxes are several $\sigma$ below those of AKARI/WISE. The same is true for the observed mm fluxes, as the JCMT and IRAM measurements are discrepant by several $\sigma$. This suggests mild variability in the disk, consistent with what is observed in the H$\alpha$ emission. The VLA and JVLA measurements are in very good agreement, except for the 3.6\,cm VLA upper limit which lies slightly below the JVLA detection.

The results of the SED modeling of $\gamma$~Cas are plotted in Fig.~\ref{gcas}. The disk contribution to the observed SED is apparent already at visual wavelengths, therefore we use only the UV spectrum to constrain the central star parameters $R_\text{p}$ and $L$. Owing to the high density of the disk, the observed fluxes from optical to far-IR were used to determine the disk parameters $\rho_0$ and $n$. The overall model fit to the observed SED from UV to far-IR region is satisfactory with the exception of the visual and near-IR fluxes, which are overestimated by the model.

The result from the modeling of the radio SED is that the disk is clearly truncated, similar to the previous targets. The disk size best reproducing the radio data is $35^{+5}_{-5}$\,$R_\text{e}$. However, the observed slope of the radio SED is again flatter than the slope of the truncated model.

   \begin{figure*}[t!]
   \centering
   \includegraphics[width=\hsize]{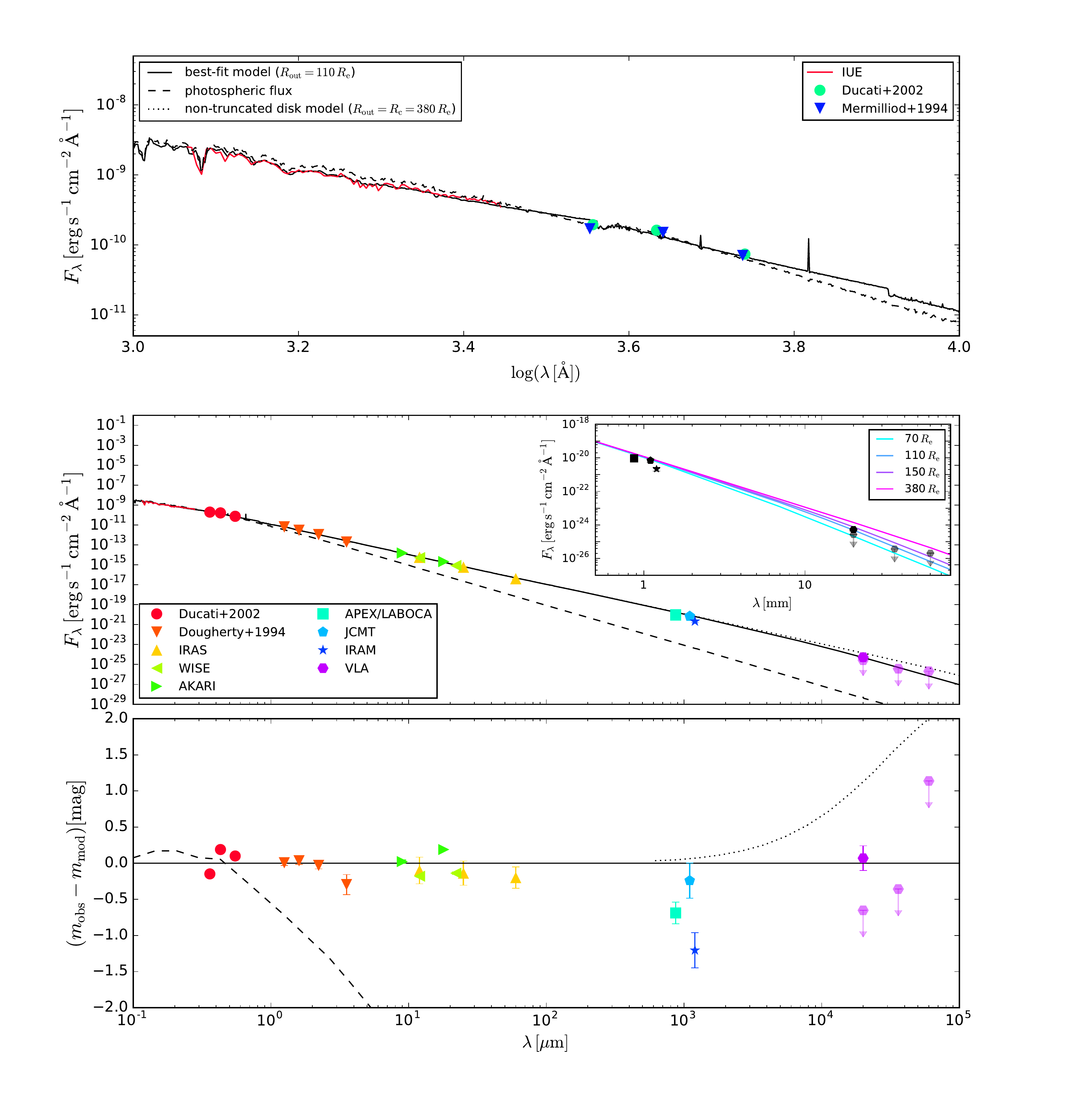}
      \caption{As in Fig.~1, but for $\beta$ Mon A.
              }
         \label{bmon}
   \end{figure*}

\subsection{$\beta$ Mon A (HD 45725; HR 2356)}

$\beta$ Mon A is a component of a visual multiple system, with the B and C components separated by 7.1 and 10 arcsec, respectively \citep{taylor}. At such distances the companions cannot have any tidal influence on the disk of the A component \citep{rivinius2006}. The star has been classified as a B3Ve star \citep{lesh, hoffleit}, B4Ve-shell star \citep{slettebak, rivinius2006}, and a B2III star \citep{maranon}. Strong V/R variations with a period of 12.5 years have been observed in the past \citep[1930--1960,][]{taylor}.

The shell nature of the star restricts the inclination angle to be close to edge-on. Unfortunately no interferometric studies of this star have been published \citep[although interferometric measurements do exist and are public -][]{ohana}; the only reference for the inclination is 67{\degree} \citep{fremat}. We adopt an inclination of 70{\degree} for our model.

There are 48 available spectra taken in four observing seasons between MJD~51000 and 56000. The narrow circumstellar line \ion{Fe}{II}\,5169 shows small RV variations with an amplitude of approximately 3\,km/s. Again, however, it is not clear whether this variation is due to binarity or changes in the disk, since $\beta$~Mon~A is a V/R variable Be star.

No HPOL data are available for $\beta$~Mon~A. Judging from the line emission, the disk is generally strong, with values of $I/I_\mathrm{c}$ between 5 and 6. It also underwent a weak V/R cyclic variability in these years. The spectra do show IR Ca triplet emission on a similar level to that of $\gamma$\,Cas.

The disk properties as shown by the observed SED seem to have been relatively stable in the last decades, with only the AKARI measurements being slightly discrepant with respect to those from IRAS and WISE. The sub-mm/mm measurements indicate a slight variability, similar to the previous targets. New JVLA observations are not available for this star and the data set at cm wavelengths consists only of a single VLA detection at 2\,cm and three upper limit measurements at 2, 3.6, and 6\,cm. The upper limit at 2\,cm lies a few $\sigma$ below the detection, suggesting a mild variability of the radio fluxes. 

The results of the SED modeling of $\beta$~Mon~A are plotted in Fig.~\ref{bmon}. $\beta$ Mon A is a shell star and the disk influence is already clear in the UV spectrum, which is slightly dimmed in comparison with a purely photospheric spectrum. The disk of $\beta$ Mon A is dense and significant excess radiation is present already at visual wavelengths. Only the IR part of the SED was used to search for the disk parameters, as the disk has a high density. The resulting model SED agrees with the observations reasonably well.

For $\beta$~Mon~A the signs of the disk truncation are the weakest among our targets. However, the APEX/LABOCA and JCMT fluxes lie more than 3-$\sigma$ below what is expected of a non-truncated disk. The upper limit measurements at 2 and 3.6\,cm are also inconsistent with a non-truncated disk. Nevertheless, additional radio measurements are necessary to better constrain the disk size and confirm the signs of truncation. However, even with the available data, the best-fit disk size is $110 \pm 40\,R_\mathrm{e}$, which is significantly lower that the critical radius value of $260\,R_\mathrm{e}$.


\section{Discussion}

The VDD model in its parametric form reproduces the observed SED of the studied stars generally well. The resulting values of the density exponent $n$ are in all cases lower than the canonical value of 3.5 derived for flaring isothermal VDDs in steady-state. A possible explanation for this was recently given by \citet{vieira2017}, who studied the IR SEDs of 80 Be stars and determined their $\rho_0$ and $n$ parameters using a semi-analytic model based on the pseudo-photosphere concept \citep{vieira}. The determined values of $n$ are generally not equal to 3.5, with the majority of the studied disks showing $n$ between 2.0 and 3.0. These results were proposed to be explained by the dynamical state that we see the disks in. According to the dynamical models \citep{haubois1} and assuming additional effects such as cooling by heavier elements (not taken into account in our model), the disks with $n \ge 3.5$ should correspond to disks that are in the process of formation, disks with $n$  between approximately 3.0 and 3.5 are in a steady-state, and disks with $n\lesssim3.0$ are in the process of dissipation. As presented in Fig.~7 of \citet{vieira2017}, approximately two thirds of the observed Be disks were found to be in the dissipating phase, suggesting that the time scales for disk dissipation are longer than those for disk formation.
 
\begin{figure}[]
   \centering
   \includegraphics[width=\hsize]{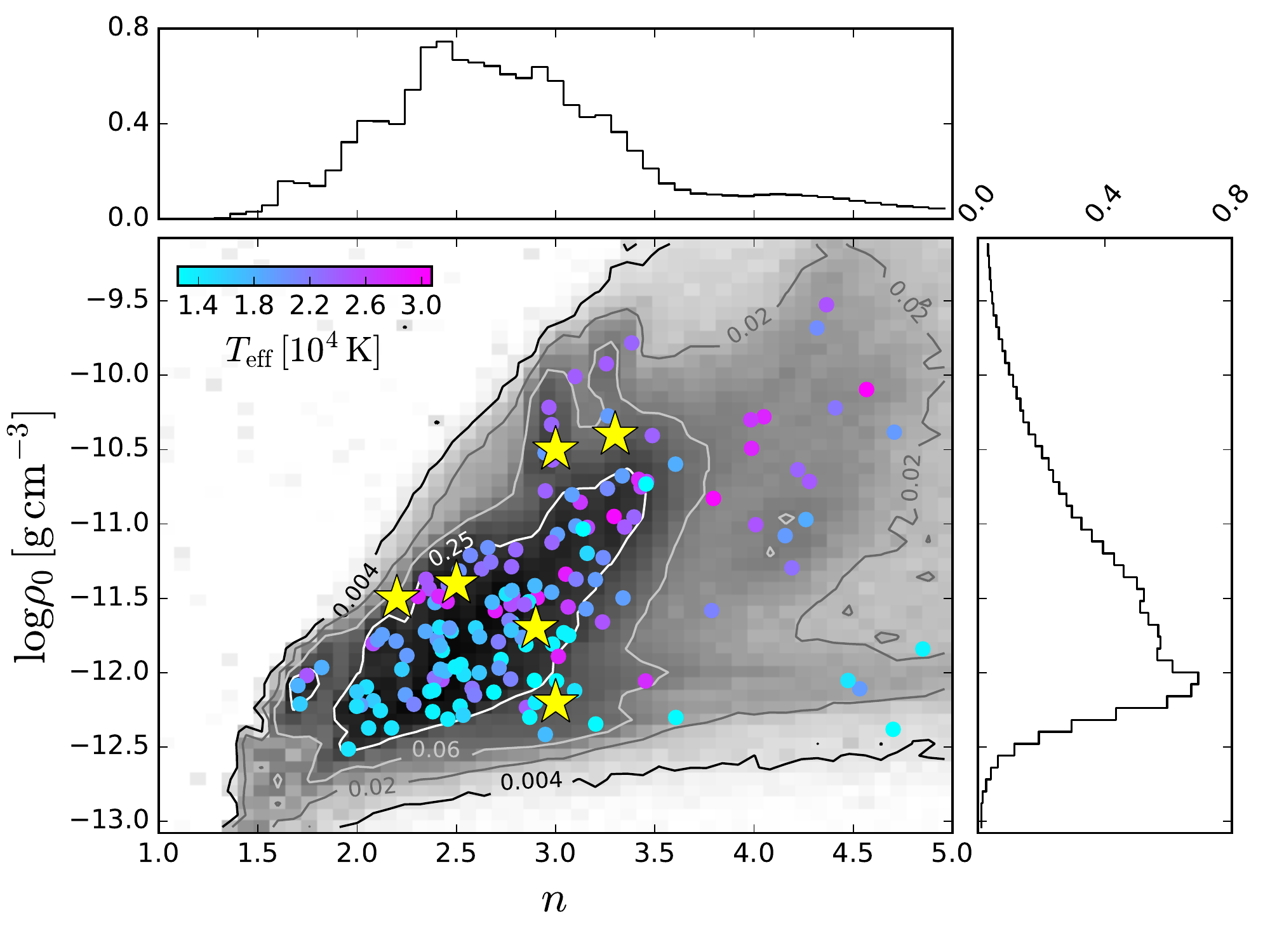}
      \caption{Results for the $\rho_0$ and $n$ distributions of the sample of Be stars studied by \citet{vieira2017}. The main panel shows the probability density (gray scale contours), while the upper and left panels show the distributions for the individual parameters. The contour values correspond to the probability density levels. Superimposed as filled circles are the results of Vieira et al. for individual stars \citep[colors indicate the stellar effective temperatures computed by][]{fremat}. Over-plotted as yellow star symbols are the results for the stars studied in this work.
              }
         \label{rodrigo}
   \end{figure}

Over-plotting the results of the present study on Fig.~7 of \citet{vieira2017} shows that the $\log{\rho_0}$ and $n$ parameter combinations determined for the stars in our sample fall within the two central contours corresponding to the probability density of 0.06 (Fig~\ref{rodrigo}). However, contrary to the study of \citet{vieira2017}, the majority of our targets seem to be in steady-state (4 out of 6). This apparent disagreement (which might not be significant given the small number of stars in our sample) is probably due to the fact that \citet{waters} favored Be stars with large infrared and radio excesses in their study. Another apparent inconsistency is that the two targets for which $n$ is well below 3.0 (EW Lac and $\psi$ Per) and should, according to \citet{vieira2017}, be in a dissipation state, also seem to have stable disks over long time-scales. We speculate that if these disks are truncated by unseen binary companions (see below), the shallower density profile might be due to the accumulation effect, which causes the disk density profile to become shallower than the steady-state value \citep{panoglou}.

An SED turndown (i.e., a steepening of the SED slope) is clearly observed between the far-IR and radio wavelengths in five out of the six studied stars. The one exception is $\beta$~Mon~A, for which the presence of an SED turndown is inconclusive.

The results of the modeling show that, to a first approximation, the observed SED turndown can be explained by assuming truncated disks. We recall that there are two physical effects that are predicted to affect the outer disk structure, that could be the cause of the SED turndown: the tidal influence from a close binary companion, or the transonic transition, that occurs when the outflow velocity of the gas in the disk becomes supersonic. 

We found that the disk truncation occurs at much smaller distances from the central star than where the supersonic point (the critical radius, $R_\text{c}$, in Table~\ref{bestfit_par}) is expected to lie. Therefore we conclude that the supersonic regime is an unlikely cause for the SED turndown observed in our sample. As mentioned in Sect.~2, the effects of the transonic transition in a typical Be disk are likely to be observable only at even longer wavelengths. The one exception is EW~Lac, for which the non-truncated disk model SED structure in the radio (dotted line in the lower panel of Fig.~\ref{ewla}) suggests that for this star (and Be stars with similarly dense disks), the effect of the transonic transition may already be observable at cm wavelengths. However, we note that the formula for the critical radius (Eq.~\ref{krticka}) is only approximate and the resulting values of $R_\mathrm{c}$ are therefore uncertain.

The disk sizes determined by our modeling represent sizes of truncated disks that best reproduce the data, but it is clear from the radio SED slope that the situation is not that simple: Some matter seems to overflow beyond the truncation radius (which is generally not equal to our parameter $R_\mathrm{out}$), from where it non-negligibly contributes to the radio SED. Investigating the role of the material past the truncation radius on the SED will be the subject of a future study.

For the previously studied star $\beta$~CMi, the presence of the companion was not confirmed at the time of the analysis of \citet{klement}. The disk truncation was nevertheless suggested as being caused by the tidal influence of an unseen binary companion, with further observable evidence, such as the Ca triplet in emission, supporting this option \citep[we refer to Sect.~5 of][]{klement}. A follow-up RV analysis of the H$\alpha$ line led to the detection of the binary, with an orbital period of 170 days and RV amplitude of 2.25\,km/s \citep{dulaney}. The value of the detected period, if a circular, coplanar orbit and a 1\,$M_\odot$ companion are assumed, indicates a semi-major axis of $\sim 50$\,$R_\text{e}$.  Based on the results of \citet{panoglou}, the truncation radius is expected to lie near the 3:1 resonance with the binary orbit, corresponding to $\sim 25$\,$R_\text{e}$. The result that the disk is truncated at $\sim 40$\,$R_\text{e}$ rather favors the 3:2 resonance with the orbit as the location of the truncation radius for the case of $\beta$~CMi.

For the other confirmed binary in our sample, $\gamma$~Cas, the semi-major axis of the orbit is $a\simeq 1.64$ au \citep{nemravova2012}, corresponding to $38\,R_\mathrm{e}$ of our model. However, here the comparison with our best-fit disk size is difficult, since the slope of the radio SED is not reproduced well. This means that if, for example, the cm data were ignored, the mm measurements would point to a much smaller disk size than was derived from the whole radio data set. This is also the case for the remaining objects with new JVLA data ($\eta$~Tau, EW~Lac, and $\psi$~Per).

In conclusion, modeling of the radio SED offers the possibility to indirectly detect previously unknown companions of Be stars. A systematic search for RV periods in stars which show truncation effects (especially shell stars), but for which no companion was reported, is necessary to confirm that it is indeed binary companions that cause the SED turndown by truncating the disk, and not any other so-far undetected mechanism operating in Be star disks. We performed a search for RV variations in the available spectra of the unconfirmed binaries in the sample. However, the data set was limited and the RV amplitudes caused by low-mass companions are likely of the same order as the value 2.25\,km/s derived for $\beta$~CMi for late-type Be stars, and even smaller for earlier types (since the mass ratio is higher). Detection of such RV amplitudes would probably require dedicated high-resolution spectroscopic campaigns and careful analysis similar to the one performed by \citet{dulaney}. If the companions are confirmed and are found to be sdO/sdB stars, this may establish a common evolutionary path for Be stars and help confirm a possible cause for Be stars being fast rotators. 

We note that the original sample of \citet{waters} was biased towards stars with the flattest spectra, that is, with the least pronounced SED turndowns. The reason is that only stars detected at cm wavelengths were analyzed, while those not detected likely have even more pronounced SED turndowns and therefore stronger signs of truncation.

\section{Conclusions}

The main conclusions of this work may be summarized as follows:

\begin{itemize}
\item The predictions of the VDD model agree very well with the observed SEDs covering the wavelength interval from the UV to the far-IR. This further establishes the VDD model as the best model so far to explain the structure of Be disks.
\item Comparison of our results with the ones of \citet{vieira2017} shows that our sample has a much higher prevalence of steeper density profiles ($n\sim3$, typically associated with disks in steady-state) than shallower profiles ($n\sim2$, usually found in Be stars whose disks are dissipating). However, for these latter cases (namely EW~Lac and $\psi$~Per) there is evidence that the disks have remained stable over the last decades, suggesting that the shallow density profile may be caused by the accumulation effect due to the unseen companions \citep{panoglou} rather than due to the disks being in a dissipation phase.
\item A clear steepening of the spectral slope at far-IR to radio wavelengths is observed in the whole sample with the exception of $\beta$~Mon~A, for which the measurements are not conclusive. The effect is present both for confirmed binaries ($\beta$~CMi, $\gamma$~Cas) as well as for stars for which no companion has yet been reported. 
\item We propose that the SED turndown is caused by truncation of the disk. The radio data sets are best reproduced by disks truncated at distances of approximately 30--150 stellar radii from the central star. The most plausible explanation for the disk truncation is the presence of (unseen) binary companions, tidally influencing the outer disk. 
\item Our simple model for truncation, in which the disk is abruptly cut at the truncation radius, cannot explain in detail the shape of the radio SED. This is in line with hydrodynamical simulations that predict that the disk material overflows past the truncation radius. In a future study we will investigate whether or not this extra material can explain the discrepancies found in this work.
\item Comparison of our results with the only previous determination of the disk size in radio \citep[][$\psi$~Per]{dougherty_nature} revealed unexpected inconsistencies. Our model that best fits the radio flux densities is too small to have been resolved by the VLA. Interestingly, it is the model with the disk size of 500\,$R_\text{e}$ that shows good agreement with the observed shape of the visibility curve. However, this model clearly overestimates the observed flux density.

\end{itemize}

\begin{acknowledgements}

We acknowledge our recently deceased colleague Stanislav \v{S}tefl as being the one who came up with the original idea for this project.

R.K. would like to acknowledge the kind help of Dietrich Baade in the final stages of preparation of the paper.

The research of R.K. was supported by grant project number 1808214 of the Charles University Grant Agency (GA UK). 

A.C.C. acknowledges support from CNPq (grant 307594/2015-7) and FAPESP (grant 2015/17967-7).

R.G.V. acknowledges support from  FAPESP (grant 2012/20364-4).

D.M.F. acknowledges support from CNPq (grant 200829/2015-7) and FAPESP (grant 2016/16844-1).

This work has made use of the computing facilities of the Laboratory of Astroinformatics (IAG/USP, NAT/Unicsul), whose purchase was made possible by the Brazilian agency FAPESP (grant 2009/54006-4) and the INCT-A.

This publication is based on data acquired with the Atacama Pathfinder Experiment (APEX). APEX is a collaboration between the Max-Planck-Institut fur Radioastronomie, the European Southern Observatory, and the Onsala Space Observatory

The National Radio Astronomy Observatory is operated by Associated Universities, Inc., under cooperative agreement with the National Science Foundation.

This publication makes use of VOSA, developed under the Spanish Virtual Observatory project supported from the Spanish MICINN through grant AyA2011-24052.

This publication makes use of data products from the Wide-field Infrared Survey Explorer, which is a joint project of the University of California, Los Angeles, and the Jet Propulsion Laboratory/California Institute of Technology, funded by the National Aeronautics and Space Administration.

The VLA data presented here were obtained as part of program VLA/10B-143 (AI141). LDM acknowledges support from an award from the National Science Foundation (AST-1516106).

\end{acknowledgements}

%
%

\bibliographystyle{aa} 
\bibliography{biblio.bib} 

\end{document}